Accepted Version

# Designing cobalt-free face-centered cubic high-entropy alloys: a strategy using d-orbital energy level

Yulin Li[a], Artur Olejarz[a], Łukasz Kurpaska[a], Eryang Lu[b], Mikko J. Alava[a,c], Hyoung Seop Kim[d,e,f], Wenyi Huo[a,*]

a- NOMATEN Centre of Excellence, National Centre for Nuclear Research, Otwock, 05-400, Poland

b- Department of Physics and Helsinki Institute of Physics, University of Helsinki, P.O. Box 43, FI-00014, Helsinki, Finland

c- Department of Applied Physics, Aalto University, P.O. Box 11000, Aalto, FI-00076, Espoo, Finland

d- Department of Materials Science and Engineering, Pohang University of Science & Technology (POSTECH), Pohang, 37673, South Korea

e- Advanced Institute for Materials Research (WPI-AIMR), Tohoku University, Sendai, 980-8577, Japan

f- Institute for Convergence Research and Education in Advanced Technology, Yonsei University, Seoul, 03722, South Korea

***Corresponding author:** Wenyi Huo, E-mail: wenyi.huo@ncbj.gov.pl.

## Abstract

High-entropy alloys (HEAs) are promising materials for high-temperature structural applications such as nuclear reactors due to their outstanding mechanical properties and thermal stability. Instead of the trial-and-error method, it is efficient to design and prepare single-phase face-centered cubic (FCC) structured HEAs using semi-empirical phase formation rules. However, almost all of phase formation rules were proposed without taking into account the cobalt-free situation. The HEAs containing cobalt are unsuitable for nuclear applications because of the long-term activation of cobalt. Here, six parameters, d-orbital energy level, valance electron concentration, entropy of mixing, enthalpy of mixing, atom size differences, and parameter of the entropy of mixing (*Ω*) were calculated to determine the solid solution phase, especially the FCC phase formation rules in cobalt-free HEAs. HEAs of 4 components were arc melted to verify the newly developed phase formation rules. The nanomechanical properties of produced HEAs were evaluated using nanoindentation. Among the six parameters, the d-orbital energy level and valance electron concentration are the critical factors that determine the FCC phase stability in cobalt-free alloys. Interestingly, the d-orbital energy level can be alone used as a benchmark for developing mechanical properties.

## 1. Introduction

The harsh service conditions of nuclear fission or fusion reactors place stringent demands on the performance of structural materials. The radiation resistance of conventional materials does not seem to satisfy the higher requirements of next-generation nuclear energy systems [1-3]. However, high-entropy alloys (HEAs) composed of multiple metallic elements in an equimolar or near equimolar ratio have attracted increasing interest due to their excellent features, such as high strength [4, 5], high ductility [6, 7], good thermal stability [8], superior wear resistance [9] and great corrosion resistance [10, 11]. It is assumed that the higher radiation resistance of HEAs resulting from their great point defect recombination ability and slower solute diffusivity under irradiation is attributed to their high configurational entropy [12]. Innovative design strategies for the chemical composition of HEAs have been proposed to replace the traditional trial-and-error approach [13-15]. However, previous studies are mainly based on Co-containing HEAs, whereas cobalt has been proven to exhibit activation under irradiation and is thus unsuitable for nuclear energy applications [15-18]. Moreover, the elimination of Co as an FCC phase stabilizer results in the instability of the FCC structure in HEAs, which then tends to convert into the BCC phase with reduced plasticity. However, the development of design strategies for Co-free HEAs with FCC structures remains a challenge.

The physical and chemical properties of HEAs are highly related to the microstructures, whereas the compositional complexity makes it difficult to predict the final structures and phases of HEAs [19]. Zhang et al. [20] summarized numerous HEAs with various compositions and calculated the corresponding enthalpy of mixing ($\Delta H_{mix}$), entropy of mixing ($\Delta S_{mix}$) and atom size differences ($\delta$). It was reported that a solid solution usually forms when $\Delta H_{mix}$ and $\Delta S_{mix}$ are in the ranges from -15 to 5 $kJ \cdot mol^{-1}$ and from 12 to 17.5 $J \cdot K^{-1} \cdot mol^{-1}$, respectively, and when $\delta$ is less than 6.5%. Another parameter, valance electron concentration (*VEC*), is also used to characterize the phases of HEAs. Researchers [21] revealed that a higher *VEC* value ($\geq 8$) facilitates the stability of the face centered cubic (FCC) phase and that the body centered cubic (BCC) phase is more stable at a lower *VEC* value (< 6.87). Lu et al. [22] were the first to apply the d-orbital energy level ($\overline{Md}$) parameter for predicting phases in HEAs. It is found that topological close-packed (TCP) phase is more stable when $\overline{Md} > 1.09$. Unfortunately, there is no unified standard for the criteria for the phase prediction of HEAs, especially the composition design rules of HEAs for radiation resistant applications, which are still lacking.

In the present work, the parameters of various reported multicomponent alloy systems are calculated carefully, including $\overline{Md}$, *VEC*, $\Delta S_{mix}$, $\Delta H_{mix}$, $\Omega$ and $\delta$. The solid solution formation rules, in particular the FCC formation rules, for Co-free HEAs were proposed in this study, and they were verified by $Co_{25}Cr_{25}Fe_{25}Ni_{25}$, $Co_{20}Cr_{20}Fe_{20}Mn_{20}Ni_{20}$, and $V_{10}Co_{10}Cr_{15}Fe_{35}Mn_{5}Ni_{25}$ alloys and a newly developed $V_{10}Cr_{20}Fe_{30}Mn_{10}Ni_{30}$ alloy.

## 2. Methods

### 2.1 Data collection

The collection of multicomponent systems was carried out using reported literature [15, 18-20, 23-63]. The alloy systems collected were all cast or subjected only to a simple homogenization treatment of water quenching, which involved as-cast and as-homogenized states. According to reports on the aging behavior of HEAs, some alloys undergo phase transformation and phase precipitation during certain aging treatments [37, 64-66]. However, investigating into whether the collected alloy systems are capable of forming solid solution phase, particularly a single-phase FCC phase, is important in this work. Therefore, the question of whether the alloy remains in a stable phase after aging treatment is not the focus of research. The definitions of $\overline{Md}$, *VEC*, $\Delta S_{mix}$, $\Delta H_{mix}$, $\Omega$ and $\delta$ are explained in the following section, and the corresponding phases as well as these parameters of various Co-containing multicomponent alloy systems are listed in Table A1. A series of Co-free multicomponent alloy systems and their parameters serving as references are also shown in Table A2. The solid solution phase is defined as a single FCC/BCC phase or a mixed FCC+BCC phase without any intermetallic phase in HEAs. HEAs with only solid solution phases or a mixture of solid solution and intermetallic phases fabricated by casting and homogenization are considered in this work.

## 2.2 Definition of parameters

$\overline{Md}$ is widely recognized as reflecting the metallurgical properties of transition group metal elements and was proposed as [22]:

$$\overline{Md} = \sum_{i=1}^{n} c_i (Md)_i \quad (1),$$

where $c_i$ is the atomic percentage of the $i$th element and $(Md)_i$ is the d-orbital energy level of the $i$th element in the $M$ element centered cluster in the $i$-$M$ binary alloy. The value of $(Md)_i$ in this study, which is consistent with Lu's work [22], was calculated based on an assumed single FCC $Ni_3Al$ structure, which can be found in the literature [67]. Another d-electron-related parameter, *VEC*, was also utilized. The *VEC* has been successfully used to determine the stability of solid solution phases and intermetallic phases in HEAs [20, 68, 69], which was defined by [20, 70]:

$$VEC = \sum_{i=1}^{n} c_i (VEC)_i \quad (2),$$

where $(VEC)_i$ is the *VEC* for the $i$th element, which is listed in Ref. [69]. Another four extensively used parameters for phase prediction are adopted in addition to the two d-electron related parameters mentioned above: $\Delta H_{mix}$, $\Delta S_{mix}$, $\Omega$ and $\delta$. They were defined by Eqs. 3-6, respectively [19, 58, 71]:

$$\Delta H_{mix} = \sum_{i=1,i\neq j}^{n} \Omega_{ij} c_i c_j \quad (3),$$

where $\Omega = 4\Delta H_{AB}^{mix}$ is an interaction parameter between the $i$th and $j$th elements in the *i* - *j* binary liquid alloy that can be found in Ref. [72]; and

$$\Delta S_{mix} = -R \sum_{i=1}^{n} c_i \ln c_i \quad (4),$$

where R is the ideal gas constant 8.314 $J\cdot K^{-1}\cdot mol^{-1}$; and

$$\Omega = \frac{T_m \Delta S_{mix}}{|\Delta H_{mix}|} \quad (5),$$

where $T_m$ is the melting temperature of the multicomponent alloy. $\Omega$ was proposed by Zhang et al. to determine the solid solution phase stability [71], and

$$\delta = 100 \times \sqrt{\sum_{i=1,i\neq j}^{n} c_i (1 - r_i/\bar{r})^2} \quad (6),$$

where $r_i$ is the atomic radius of the $i$th element and $\bar{r} = \sum_{i=1}^{n} c_i r_i$ is the average atomic radius.

## 2.3 Experimental details

$Co_{25}Cr_{25}Fe_{25}Ni_{25}$ and $Co_{20}Cr_{20}Fe_{20}Mn_{20}Ni_{20}$ are both typical alloy systems with stable single FCC phases that have been extensively studied [73, 74]. The $V_{10}Co_{10}Cr_{15}Fe_{35}Mn_5Ni_{25}$ alloy, which has a more complex alloy system with more constituent elements and a stable single FCC phase, was also selected [75]. Vanadium in certain HEAs shows strong miscibility with other transition elements, and can form a single-phase solid solution phase in a wide range of binary phase diagrams [75]. The atomic size of V is slightly larger than that of the other atoms in certain alloy systems. The lattice distortion caused by V can result in solid solution strengthening. The selection of these three representative alloy systems as references, which exhibit a stable single FCC phase, facilitates the fabrication of Co-containing HEAs with a single FCC phase, thereby enabling a comparative assessment of phase stability with Co-free HEAs. A newly designed $V_{10}Cr_{20}Fe_{30}Mn_{10}Ni_{30}$ alloy has been modified to exclude the element Co and to stabilize the values of semi-empirical parameters within the established range for a single FCC phase, based on the $V_{10}Co_{10}Cr_{15}Fe_{35}Mn_5Ni_{25}$ alloy. The exclusion of Co is intended to satisfy the requirements of nuclear applications.

The HEAs studied in this work, i.e., $Co_{25}Cr_{25}Fe_{25}Ni_{25}$, $Co_{20}Cr_{20}Fe_{20}Mn_{20}Ni_{20}$, $V_{10}Co_{10}Cr_{15}Fe_{35}Mn_{5}Ni_{25}$, and $V_{10}Cr_{20}Fe_{30}Mn_{10}Ni_{30}$, were prepared by arc-melting in a high-purity argon atmosphere. The ingots, consisting of high purity metals, i.e., Co (≥ 99.5%), Cr (≥ 99.5%), Fe (≥ 99.99%), Mn (≥ 99.99%), Ni (≥ 99.97%) and V (≥ 99.7%), were flipped and remelted 6 times to ensure chemical homogeneity. The as-cast alloys were homogenized at 1200 °C for 4 h in an argon atmosphere, followed by water quenching. The crystal structure of the samples was characterized by using D8 Discover X-ray diffraction (XRD) with Cu $K\alpha$ radiation. A voltage of 40 kV, a current of 40 mA and a scan speed of $2°\cdot s^{-1}$ were used. Nanoindentation tests were performed on an MML Nano Test Vantage using a Berkovich diamond tip indenter at ambient temperature. To ensure the accuracy and reliability of the results, a 7 × 7 array of indentations was made for each case. Tests were conducted at a constant strain rate of $5\ mN\cdot s^{-1}$ and a maximum load of 150 mN.

## 3. Results

### 3.1 Effects of individual parameters

The effects of parameters $\delta$, $VEC$, $\overline{Md}$, $\Omega$, $\Delta S_{mix}$ and $\Delta H_{mix}$ on multicomponent alloys can be clearly shown in Fig. 1, which is plotted by summarizing the data in Tables 1 and 2. The alloys with a single BCC/FCC phase or a BCC and FCC mixed phase are denoted as the solid solution phase (SS) in Fig. 1. From Fig. 1a, it can be seen that the range of $\delta$ for Co-free alloys is shorter than the counterparts of Co-containing alloys, and $\delta$ values are distributed in the ranges from 2.53% to 7.08% and from 0 to 6.6%, respectively. Conversely, the ranges of $VEC$ and $\overline{Md}$ for Co-free alloys are slightly wider, as shown in Figs. 1b and c. The solid solution phase is more stable in Co-free alloys when $VEC$ and $\overline{Md}$ are in the ranges from 6.46 to 9.00 and from 0.787 to 1.260, respectively. The decrease in VEC and increase in $\overline{Md}$ result from the substitution of elements with lower $VEC$ values and higher $\overline{Md}$ values, such as Nb, Ti, Zr and Ta [67, 69], for Co. The individual parameters $\Omega$ and $\Delta H_{mix}$ seem to have no significant effect on the formation of the solid solution phase between Co-free alloys and Co-containing alloys, as shown in Fig. 1d and f. $\Delta S_{mix}$, in Fig. 1e, shows a weak effect on the solid solution phase stability between the two types of alloys. A solid solution forms in Co-free alloys when $\Delta S_{mix}$ is low; in contrast, the solid solution phase of Co-containing alloys tends to form when $\Delta S_{mix}$ is slightly higher. It is noteworthy that in most cases, the distribution of the solid solution and intermetallic mixed phase almost coincides with the solid solution phase, which is consistent with other reports [70, 71].

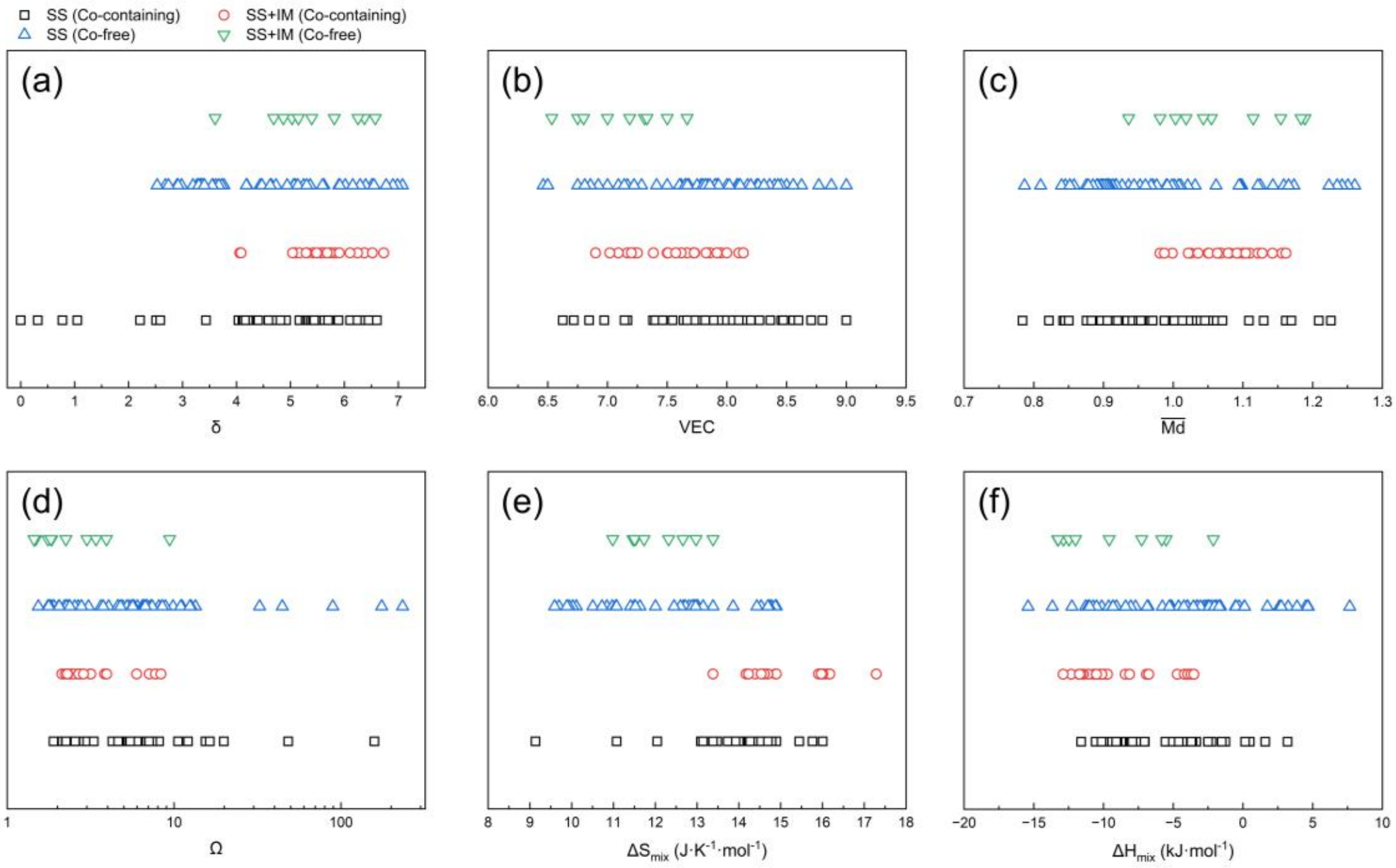

**Fig. 1** Distributions of parameters *δ, VEC,* $\overline{Md}$*, Ω,* $\Delta S_{mix}$ and $\Delta H_{mix}$ for HEAs. The symbol, □, represents SS phase forming Co-containing alloys, ○ represents SS and intermetallic mixed phase forming Co-containing alloys, △ represents SS phase forming Co-free alloys, and ▽ represents SS and intermetallic mixed phase forming Co-free alloys.

### 3.2 Collective effect of parameters

According to the Hume-Rothery rule, one of the necessary conditions to form solid solution phases is to have relatively small atomic size differences [76]. To better explore the comprehensive effect of multiple factors on phase stability, five figures plotted by *δ* superimposed with *VEC,* $\overline{Md}$*, Ω,* $\Delta S_{mix}$ and $\Delta H_{mix}$ are shown in Fig. 2. The parameters of Co-free HEAs indicate a different distribution from Co-containing HEAs.

From Fig. 2a, it is revealed that *VEC* has a negative correlation with $\delta$ and that the solid solution phase exhibits a nearly linear decrease in *VEC* with increasing $\delta$. The fitting curves of the two solid solution phases show that the *VEC* of Co-containing alloys varies more drastically with $\delta$ than that of Co-free alloys. The fitting curves of the solid solution phase for Co-containing and Co-free alloys are defined as $y = -0.06x^2 + 0.15x + 8.66$ and $y = -0.06x^2 + 0.27x + 7.97$, respectively. This demonstrates that the solid solution phases of Co-free alloys have lower *VEC* values when $\delta < 5.75$ % and higher *VEC* values than Co-containing alloys when $\delta$ exceeds 5.65%. The parameter $\overline{Md}$ shows an opposite fitting trend to *VEC*, as shown in Fig. 2b. Compared to *VEC*, the distribution of the phases in the $\overline{Md}$ - $\delta$ map is more convergent. The corresponding fitting curves for Co-containing and Co-free alloys are defined as $y = 0.01x^2 - 0.03x + 0.85$ and $y = 0.02x^2 - 0.11x + 1.02$, respectively. In contrast to *VEC*, the $\overline{Md}$ values of Co-free alloys are higher than those of Co-containing alloys when $\delta$ is less than 3.98%. The $\overline{Md}$ values increase more rapidly with $\delta$ when $\delta$ is larger than 3.98%. As mentioned above, it is difficult to dedicate the distribution difference of the solid solution phase between Co-free and Co-containing HEAs by parameters $\Omega$ and $\Delta H_{mix}$ from Figs. 2c and e. Nevertheless, the clear ranges of $\delta \leq 7.08\%$, $\Omega \geq 1.54$ and $-15.40 \leq \Delta H_{mix} \leq 7.66$ kJ·mol$^{-1}$ for solid solution phase formation in Co-free HEAs can be verified. For $\Delta S_{mix}$ in Fig. 2d, the solid solution phase in Co-free HEAs stabilized with smaller values, from 9.59 to 14.90 J·K$^{-1}$·mol$^{-1}$, than that of Co-containing HEAs. When $\Delta S_{mix} > 14.90$ J·K$^{-1}$·mol$^{-1}$, the high entropy of the mixing value facilitates the formation of a solid solution phase and intermetallic phase in Co-containing HEAs.

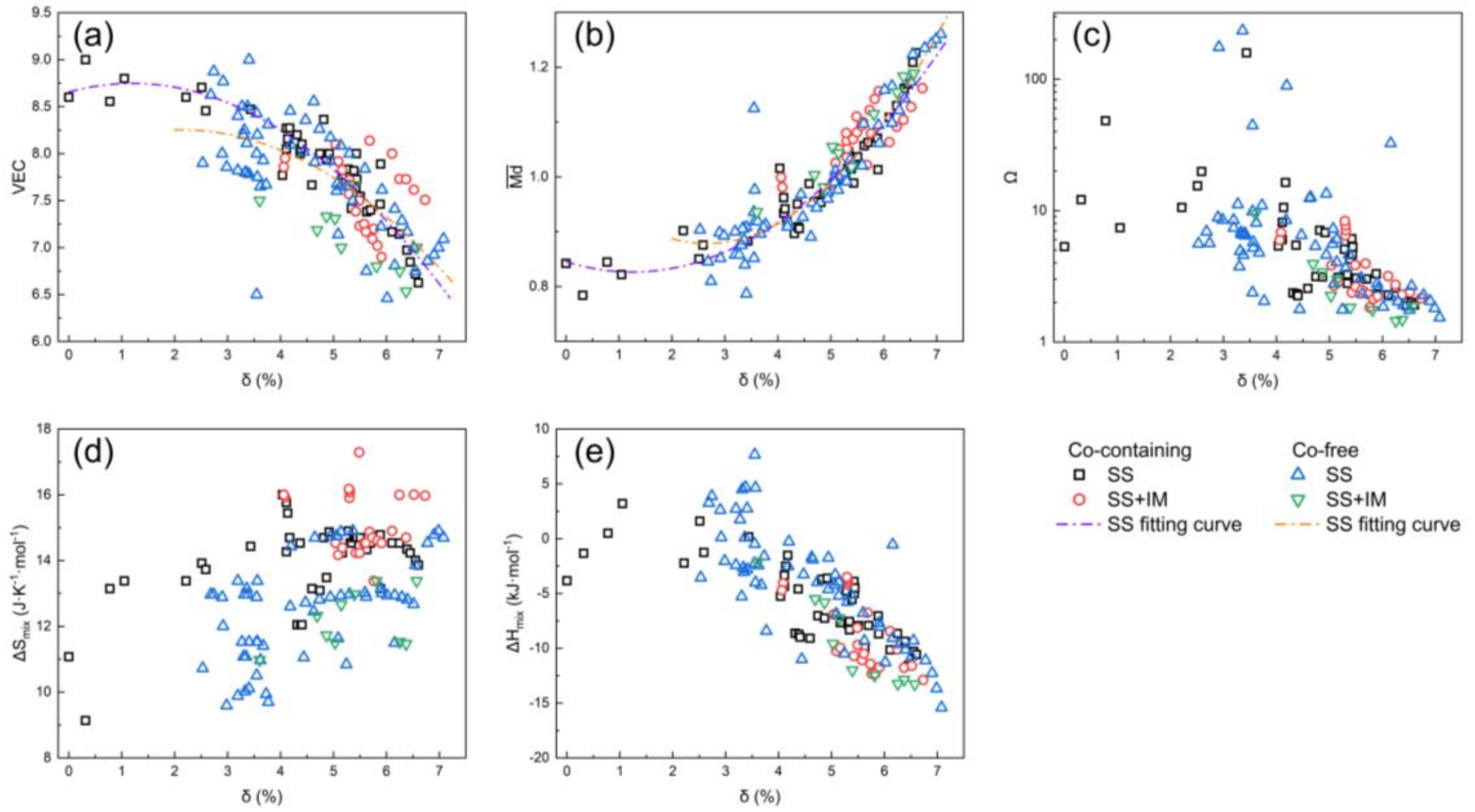

**Fig. 2** Distributions of $\delta$ superimposed with parameters (a)*VEC*, (b) $\overline{Md}$, (c) $\Omega$, (d) $\Delta S_{mix}$ and (e) $\Delta H_{mix}$. The purple and orange dash-dotted lines in (a) and (b) are fit curves of SS phases in Co-containing and Co-free alloys, respectively.

The similar trends in *VEC* and $\overline{Md}$ in Fig. 2 indicate that a correlation exists between these two factors. A series of combinations among *VEC* and $\overline{Md}$, $\overline{Md}$ and $\Delta H_{mix}$ as well as *VEC* and $\Delta H_{mix}$ that have linear correlation are plotted in Fig. 3. A significant negative linear correlation between *VEC* and $\overline{Md}$ is exhibited in Fig. 3a. A fitting line of the SS phase with a steeper slope can be observed in Co-free alloys than in Co-containing alloys. When $\overline{Md}$ < 1.08, a higher *VEC* is needed to form stable SS phases in Co-free alloys, and when $\overline{Md}$ > 1.08, the formation of SS phases in Co-free alloys requires less *VEC* than in Co-containing alloys. Interestingly, most of the SS and intermetallic mixed phases in Co-containing and Co-free alloys are distributed above and below the fitting line, respectively. $\Delta H_{mix}$ also has a negative correlation with $\overline{Md}$, as shown in Fig. 3b. The $\Delta H_{mix}$ needed for SS phase formation is higher overall for Co-free alloys at the same $\overline{Md}$. In contrast to the relationships between $\overline{Md}$ and VEC, and $\overline{Md}$ and $\Delta H_{mix}$, two near-parallel fitting lines are plotted in the $\Delta H_{mix}$ - *VEC* map, as shown in Fig. 3c. The solid solution phase stability of Co-free and Co-containing alloys has a similar sensitivity to $\Delta H_{mix}$ at the same *VEC*. Although there is overlap between the mixed phase and solid solution phase regions, the mixed phase remains stable in the regions of *VEC* < 8.14 and $\overline{Md}$ > 0.981. This region is suggested to be avoided when designing solid solution-phase HEAs.

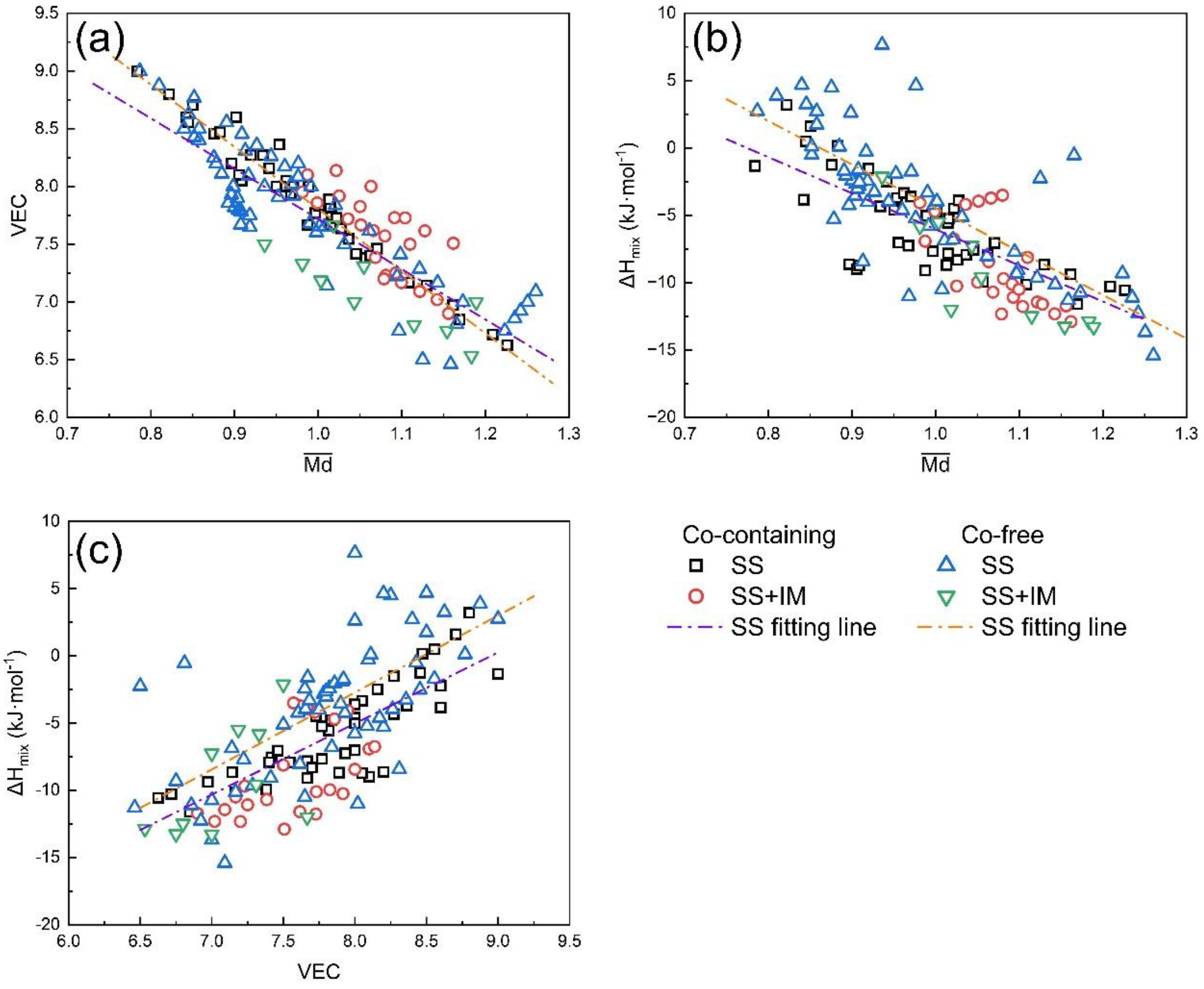

**Fig. 3** Binary factor map among (a) $\overline{Md}$ and *VEC*, (b) $\overline{Md}$ and $\Omega$, and (c) $\overline{Md}$ and $\Delta H_{mix}$ on SS phase stability in Co-containing and Co-free alloys.

### 3.3 FCC phase formation rules for Co-free alloys

HEAs with FCC structures have garnered significant attention as viable reactor structural materials due to their exceptional thermomechanical properties [77-79]. Therefore, the SS phases are classified as BCC, FCC, and mixed phases to derive the FCC phase formation rules about Co-free alloys in *VEC* - $\overline{Md}$, $\Delta H_{mix}$ - $\Delta S_{mix}$, and $\Omega$ - $\delta$ maps, as presented in Fig. 4. The smoother slope of the FCC fitting curve shown in Fig. 4a for Co-free alloys indicates a reduced susceptibility to modifications in the *VEC* and $\overline{Md}$, which may contribute to enhanced FCC phase stability in these materials. The FCC phase of Co-free alloys can be stabilized when 0.787 ≤ $\overline{Md}$ ≤ 0.992 and 7.67 ≤ *VEC* ≤ 9.00. The FCC phase stable regions in Co-free alloys are smaller than those in Co-containing alloys, as indicated by the two sets of parallel dash-dotted lines in Figs. 4b and c. The stability of the FCC phase in Co-free alloys is enhanced when the values of $9.59 \leq \Delta S_{mix} \leq 13.38$ J·K$^{-1}$·mol$^{-1}$, $-11.00 \leq \Delta H_{mix} \leq 4.64$ kJ·mol$^{-1}$, $2.53\% \leq \delta \leq 5.30\%$, and $\Omega \geq 1.77$ are met.

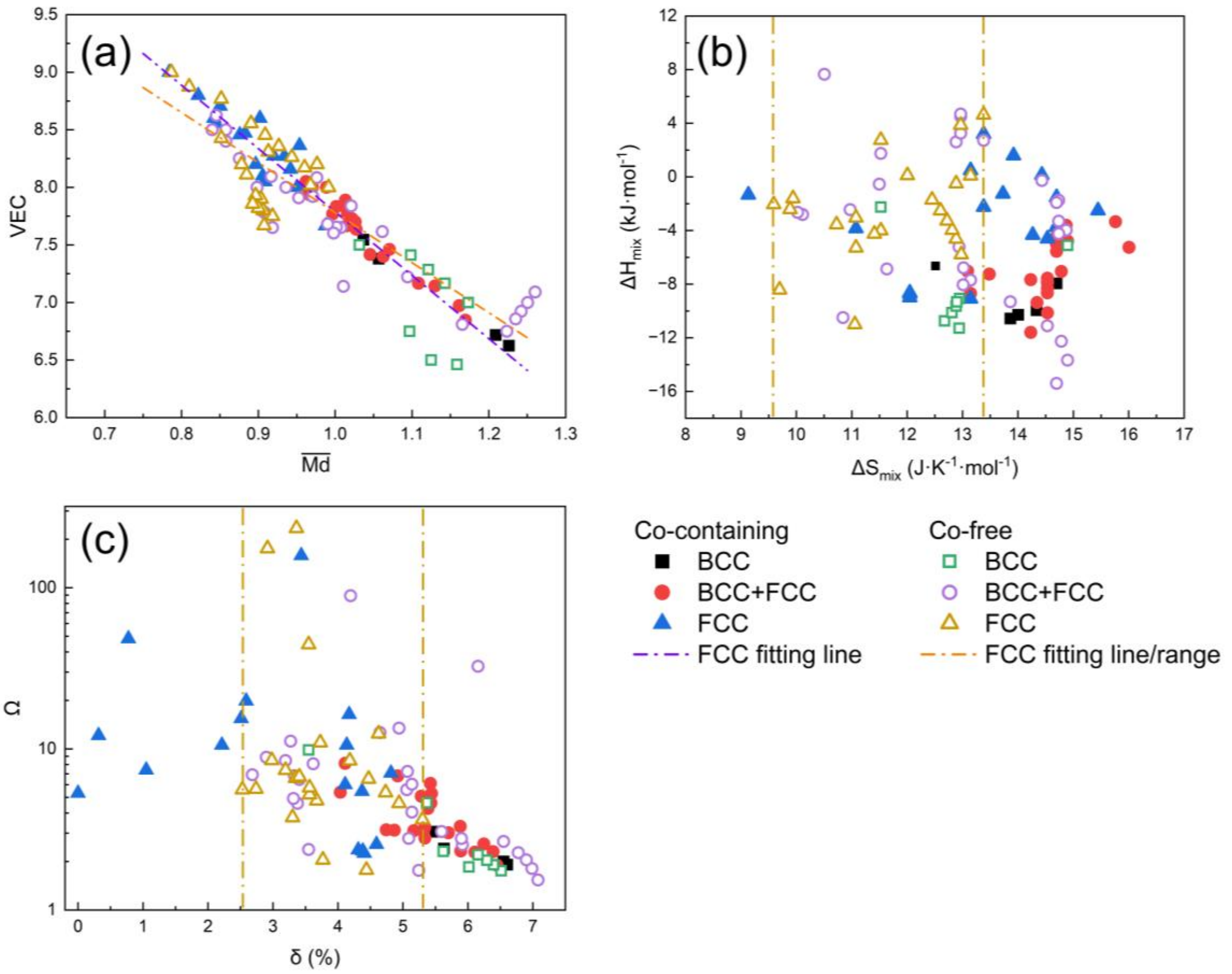

**Fig. 4** Binary factor map among (a) $\overline{Md}$ and $VEC$, (b) $\Delta S_{mix}$ and $\Delta H_{mix}$, and (c) $\delta$ and $\Omega$ on SS phase stability in Co-containing and Co-free alloys.

### 3.4 Validation of FCC phase formation rules

To verify the phase formation rules demonstrated in Figs. 2-4, three classical multicomponent systems of $Co_{25}Cr_{25}Fe_{25}Ni_{25}$, $Co_{20}Cr_{20}Fe_{20}Mn_{20}Ni_{20}$, and $V_{10}Co_{10}Cr_{15}Fe_{35}Mn_{5}Ni_{25}$ were manufactured, and a new alloy system of $V_{10}Cr_{20}Fe_{30}Mn_{10}Ni_{30}$ was developed. Fig. 5 exhibits the XRD patterns of four different HEAs subjected to homogenizing and water quenching. As shown in Fig. 5a, the Co-containing samples $Co_{25}Cr_{25}Fe_{25}Ni_{25}$, $Co_{20}Cr_{20}Fe_{20}Mn_{20}Ni_{20}$ and $V_{10}Co_{10}Cr_{15}Fe_{35}Mn_5Ni_{25}$ present only a single FCC phase. In $Co_{20}Cr_{20}Fe_{20}Mn_{20}Ni_{20}$ and $V_{10}Cr_{15}Mn_5Fe_{35}Co_{10}Ni_{25}$, the peak intensities of (200) and (111) are extremely high, indicating the formation of a strong texture in the alloy [80]. The 2θ degree of the non-FCC phase peak observed in the $V_{10}Cr_{15}Mn_5Fe_{35}Co_{10}Ni_{25}$ alloy is consistent with the degree of the BCC phase (110) peak reported in previous studies [19, 20, 44, 64], which can be confirmed as the BCC phase. The 2θ degrees of the FCC and BCC peaks and the calculated lattice constants of the FCC structures are summarized in Table 1. The change in the lattice constant due to the different compositions of the HEAs is reflected in Fig. 5b as a slight shift of the peaks. A weak peak of the BCC phase is detected in Co-free $V_{10}Cr_{20}Fe_{30}Mn_{10}Ni_{30}$. A very small volume fraction of the BCC phase forms in the main FCC phase of the Co-free alloy.

Table 1 presents the calculated values of the lattice constants, $\overline{Md}$, $VEC$, $\Delta S_{mix}$, $\Delta H_{mix}$, $\Omega$ and $\delta$ of the four HEAs. The lattice constants of the four alloy systems increase with the addition of V and the increase in Mn content after the replacement of Co. The parameters of the three classical Co-containing alloy systems $Co_{25}Cr_{25}Fe_{25}Ni_{25}$, $Co_{20}Cr_{20}Fe_{20}Mn_{20}Ni_{20}$ and $V_{10}Co_{10}Cr_{15}Fe_{35}Mn_5Ni_{25}$ are constant with the solid solution formation rules proposed by Zhang et al. [19, 71]. The parameters of the newly developed $V_{10}Cr_{20}Fe_{30}Mn_{10}Ni_{30}$ are all within the aforementioned solid solution formation rules for Co-free alloys, which are $0.787 \leq \overline{Md} \leq 1.260$, $6.46 \leq VEC \leq 9.00$, $9.59 \leq \Delta S_{mix} \leq 14.90$ J·K$^{-1}$·mol$^{-1}$, $-15.40 \leq \Delta H_{mix} \leq 7.66$ kJ·mol$^{-1}$, $\Omega \geq 1.54$ and $\delta \leq 7.08$ %. The detected peak in the BCC phase in the new alloy may be explained by the proximity of the values of $\overline{Md}$ and $VEC$ to the boundaries of the FCC phase formation rules shown in Fig. 4, as their ranges are restricted. It is suggested that $\overline{Md}$ and $VEC$ are the more critical parameters determining the formation of the FCC phase in Co-free alloys. Despite the presence of a few BCC phases, the formation of the main FCC phase supports the logic behind the FCC formation rules for Co-free alloys, which are $0.787 \leq \overline{Md} \leq 0.992$, $7.67 \leq VEC \leq 9.00$, $9.59 \leq \Delta S_{mix} \leq 13.38$ J·K$^{-1}$·mol$^{-1}$, $-11.00 \leq \Delta H_{mix} \leq 4.64$ kJ·mol$^{-1}$, $2.53\% \leq \delta \leq 5.30$ %, and $\Omega \geq 1.77$.

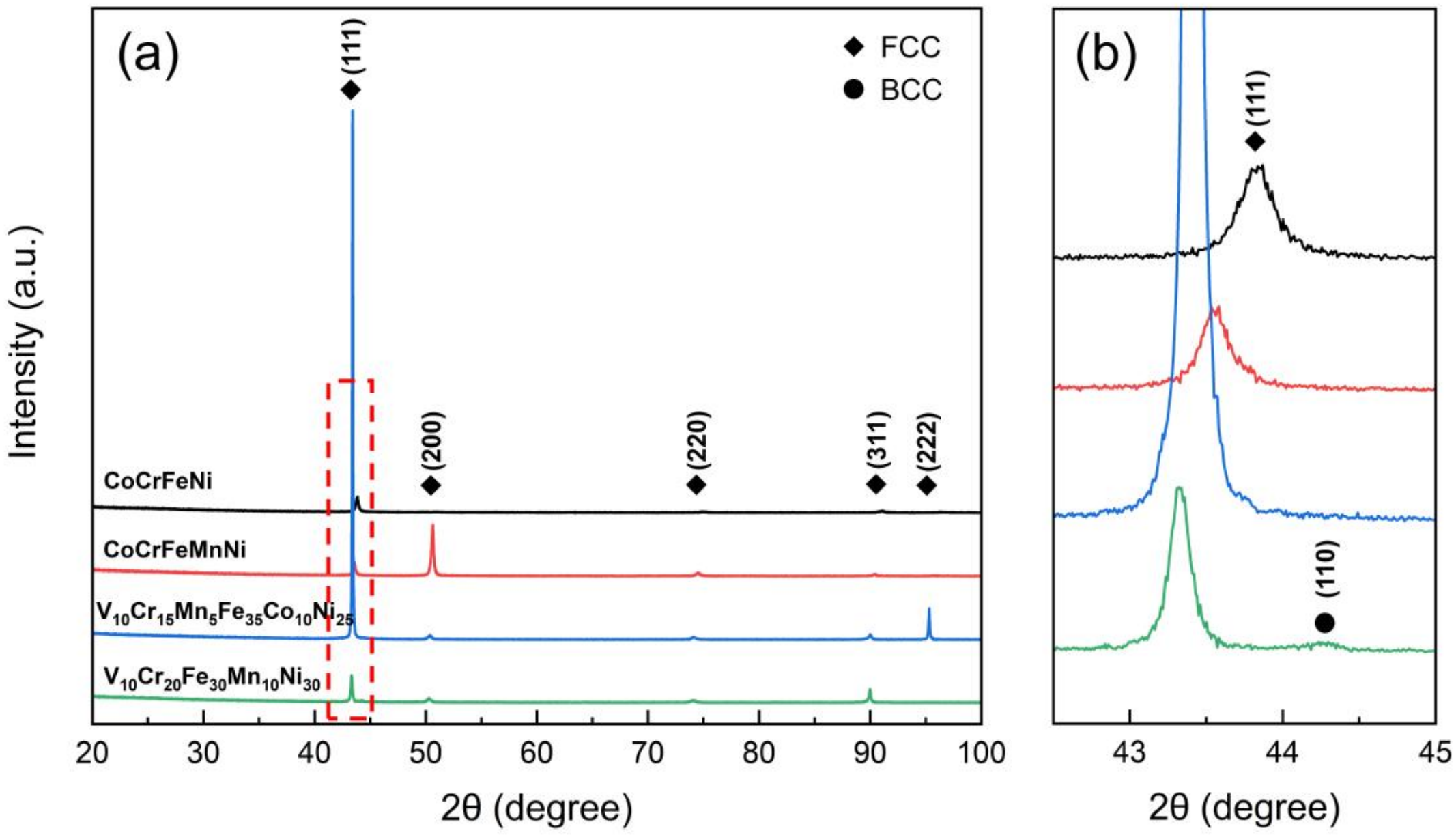

**Fig. 5** (a) XRD patterns of four HEAs, and (b) enlarged view with a 2θ from 42.5° to 45°.

**Table 1** Calculated results from XRD tests.

| Materials | FCC (°) | | | | | BCC (°) | FCC α (Å) |
|---|---|---|---|---|---|---|---|
| | 111 | 200 | 220 | 311 | 222 | 110 | |
| $Co_{25}Cr_{25}Fe_{25}Ni_{25}$ | 43.84 | 50.95 | 74.88 | 91.13 | - | - | 3.576 |
| $Co_{20}Cr_{20}Fe_{20}Mn_{20}Ni_{20}$ | 43.55 | 50.60 | 74.45 | 90.53 | - | - | 3.596 |
| $V_{10}Co_{10}Cr_{15}Fe_{35}Mn_{5}Ni_{25}$ | 43.43 | 50.36 | 74.09 | 90.01 | 95.33 | - | 3.606 |
| $V_{10}Cr_{20}Fe_{30}Mn_{10}Ni_{30}$ | 43.27 | 50.28 | 74.01 | 89.96 | - | 44.20 | 3.613 |

**Table 2** Calculated parameters of $\overline{Md}$, *VEC*, $\Delta S_{mix}$, $\Delta H_{mix}$, $\Omega$, $\delta$, and phase of four HEAs.

| Materials | $\overline{Md}$ | *VEC* | $\Delta S_{mix}$ ($J \cdot K^{-1} \cdot mol^{-1}$) | $\Delta H_{mix}$ ($kJ \cdot mol^{-1}$) | $\Omega$ | $\delta$ | Phase |
|---|---|---|---|---|---|---|---|
| $Co_{25}Cr_{25}Fe_{25}Ni_{25}$ | 0.874 | 8.25 | 11.53 | -3.75 | 5.75 | 0.32 | FCC |
| $Co_{20}Cr_{20}Fe_{20}Mn_{20}Ni_{20}$ | 0.890 | 8.00 | 13.38 | -4.16 | 5.79 | 3.27 | FCC |
| $V_{10}Co_{10}Cr_{15}Fe_{35}Mn_{5}Ni_{25}$ | 0.931 | 7.95 | 13.38 | -6.26 | 3.98 | 2.41 | FCC |
| $V_{10}Cr_{20}Fe_{30}Mn_{10}Ni_{30}$ | 0.951 | 7.80 | 12.51 | -6.64 | 3.52 | 2.86 | FCC+BCC |

### 3.5 Nanomechanical properties

The typical load–displacement (P-h) curves of the four HEAs are shown in Fig. 6a. The HEAs of varying compositions demonstrate a reduction in their maximum indentation depth when subjected to identical loads. The indentation results show that $V_{10}Cr_{20}Fe_{30}Mn_{10}Ni_{30}$ possesses the highest hardness compared to the other three alloys. The mechanical properties of HEAs, including hardness and elastic modulus, are plotted in Fig. 6b, and the specific values are summarized in Table 3. The hardness of the HEAs increases from 2.08 GPa to 2.44 GPa. Interestingly, there is no indication of a decrease in the elastic modulus with increased hardness. The new alloy possesses almost the same level of elastic modulus as classical $Co_{20}Cr_{20}Fe_{20}Mn_{20}Ni_{20}$. The nanoindentation tests show that Co-free $V_{10}Cr_{20}Fe_{30}Mn_{10}Ni_{30}$ with FCC + BCC phases exhibit high hardness with excellent elastic modulus.

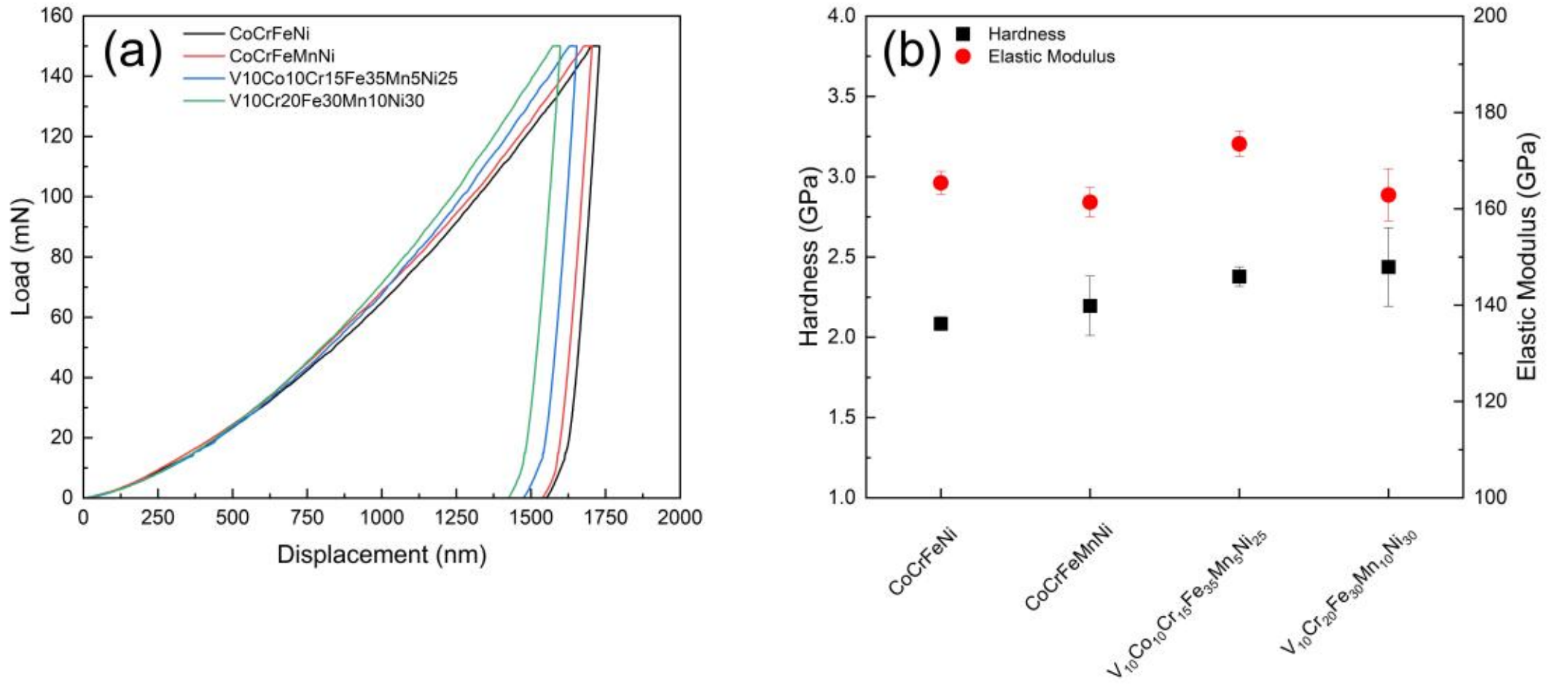

**Fig. 6** (a) Representative load–displacement curves of four HEAs during nanoindentation tests with a Berkovich indenter, and (b) the corresponding hardness and elastic modulus.

**Table 3** Nanomechanical properties of the four HEAs.

| Materials | Hardness (GPa) | Elastic modulus |
|---|---|---|
| $Co_{25}Cr_{25}Fe_{25}Ni_{25}$ | 2.08 ± 0.04 | 165.39 ± 2.40 |
| $Co_{20}Cr_{20}Fe_{20}Mn_{20}Ni_{20}$ | 2.20 ± 0.18 | 161.36 ± 3.05 |
| $V_{10}Co_{10}Cr_{15}Fe_{35}Mn_{5}Ni_{25}$ | 2.38 ± 0.06 | 173.47 ± 2.66 |
| $V_{10}Cr_{20}Fe_{30}Mn_{10}Ni_{30}$ | 2.44 ± 0.24 | 162.88 ± 5.44 |

## 4. Discussion

### 4.1 Effects of various parameters on phase formation for Co-free HEAs

The enthalpy of mixing and entropy of mixing are widely used to predict solid solution phase stability in HEAs. $\Delta H_{mix}$ represents the miscibility of various elements in the liquid alloy [71]. Theoretically, the closer $H$ is to zero, the more favorable it is for the formation of random solid solutions. However, whether in past studies or in the current research, it has been observed that the range for stabilization of the solid solution phase is broader when $\Delta H_{mix}$ is in negative values. The parameter $\Delta S_{mix}$ should be considered as well as $\Delta H_{mix}$ when predicting the phase stability. The value of $\Delta S_{mix}$ indicates the extent of confusion of the alloy system, and multicomponent systems generally possess a positive $\Delta S_{mix}$ [71]. The large $\Delta S_{mix}$ value can noticeably decrease the system's Gibbs free energy and enable the random dispersal of different elements in the lattice, thereby inhibiting the formation of ordered phases, such as B2, σ, and Laves phases [81, 82]. Guo et al. [69] proposed a smaller range of values for solid solution formation in HEAs based on the theory of Zhang et al. [19]: $-22 \le \Delta H_{mix} \le 7$ $kJ \cdot mol^{-1}$ and $11 \le \Delta S_{mix} \le 19.5$ $J \cdot K^{-1} \cdot mol^{-1}$. However, for Co-free HEAs, the value range should be more precise, which is determined as $-15.40 \le \Delta H_{mix} \le 7.66$ $kJ \cdot mol^{-1}$ and $9.59 \le \Delta S_{mix} \le 14.90$ $J \cdot K^{-1} \cdot mol^{-1}$, and the values for FCC phase formation are $-11.00 \le \Delta H_{mix} \le 4.64$ $kJ \cdot mol^{-1}$ and $9.59 \le \Delta S_{mix} \le 13.38$ $J \cdot K^{-1} \cdot mol^{-1}$ according to the present study. As indicated in Fig. 1 and Fig. 4, $\Delta S_{mix}$ plays a more significant role in differentiating between Co-containing and Co-free phases in regard to the formation rules of solid solution and FCC phases than $\Delta H_{mix}$.

The empirical parameter $\Omega$ combined effects of $T_m \Delta S_{mix}$ and $\Delta H_{mix}$ were proposed to be one of the solid solution phase criteria in 2012 [71]. When $\Omega$ fluctuates above and below 1, $T_m \Delta S_{mix}$ and $\Delta H_{mix}$ alternate in dominance. It is more likely to form a solid solution phase if the value of $\Delta S_{mix}$ is high enough to offset the positive or large negative $\Delta H_{mix}$ [71, 81]. In the current case, the solid solution phases of Co-containing and Co-free alloys have the same range of $\Omega$ values. However, a precise range of $\Omega$ can be determined. To stabilize the solid solution phase, $\Omega$ is $\ge 1.54$, while for FCC phase stabilization in Co-free alloys, *Ω* should be ≥ 1.77 in Co-free alloys. The trend can be indicated by the different preferences of $\Delta S_{mix}$ between Co-free and Co-containing alloys from Figs. 1 and 4. Another physical parameter typically accompanied by *Ω* is *δ*. Excessive *δ* causes lattice distortion, impeding atomic diffusion and hindering solid solution phase formation, ultimately leading to the production of an intermetallic phase or glassy phase [69, 71, 83]. In most reports, *δ* has a significant effect on the phase prediction of HEAs. Guo et al. [69] extended the upper limit of solid solution formation rules for *δ* in HEAs from 6.6 to 8.5%. Senkov et al. [84] reported a boundary of *δ* = 3.47% between the FCC and intermetallic phases, despite *Ω* and $\Delta H_{mix}$ being ineffective in their case. According to the present work, the FCC phase in Co-free alloys forms when *δ* ≤ 5.30%. The smaller *δ* range for FCC phase formation in Co-free alloys may be attributed to the substitution of elements with larger atomic size differences, such as V, Mo and Mn, for Co. To avoid the adverse effects of large atomic differences on the formation of FCC phases, a more stringent limit is needed for Co-free alloys.

### 4.2 Valence electron-related parameters *VEC* and $\overline{Md}$

There is a substantial association between *VEC* and $\overline{Md}$ in estimating the phase stability of HEAs, with both variables being correlated with d-electrons, as evidenced by Figs. 2, 3 and 4. When considering electron concentration, two concepts are usually considered: the average number of valence electrons per atom (*e/a*) and the number of total valence electrons including atoms filled in d-band (*VEC*) [68, 85, 86]. Due to the uncertain nature of *e/a* on transition metals, this work solely adopts *VEC* [85, 87]. This method has proven to be essential in distinguishing between the FCC and BCC phases of HEAs [44, 70, 85]. It is widely acknowledged that a high *VEC* favors the stability of the FCC phase, while the opposite is true for the BCC phase. Guo et al. [20] proposed that the FCC phase is stable at *VEC* ≥ 8, the BCC phase is stable at *VEC* ≤ 6.87 and a mixture of FCC and BCC phases forms in the middle region. Afterwards, a *VEC* threshold of approximately 7.5 was reported to distinguish between FCC and BCC phases [70]. In addition to the inherent FCC lattice of FCC stabilizing elements, such as Co, Cu and Ni, their high *VEC* values also lead to elevated *VEC* values in FCC-structured HEAs. Therefore, the threshold for stabilization of the FCC phase is lowered when Co in the HEA is replaced by Mn, V and/or Mo with lower *VEC* values. In the present research, a single FCC phase forms in Co-free alloys when the *VEC* is above 7.67.

The average value of the energy level of the d-orbital, i.e., $\overline{Md}$, is another parameter regarding d-electrons. According to the theory of Morinaga et al. [88], $\overline{Md}$ correlates with the electronegativity and atomic radius of both solute and solvent elements. The transfer of charge from elements with higher d-level energy to those with lower d-level energy is governed by the electronegativity of the atoms involved. Thus, $\overline{Md}$ shows a trend opposite to electronegativity. The energy of the d-level increases as the d-orbital radius increases, resulting in a high value of $\overline{Md}$ [89]. The $\overline{Md}$ method has been successfully used in the phase stability prediction of Ni-based [90] alloys, Cr-based alloys [91], Ti-based alloys [92] and Fe-based alloys [93]. It was found that the selection of the main elements affects the $\overline{Md}$ threshold values for the crystal structure of alloys. However, HEAs lack primary constituents and exhibit near-equimolar elemental compositions. The $\overline{Md}$ method is rarely utilized to estimate phase stability in HEAs, and it is necessary to confirm its practicability in HEAs to further develop superior performance HEAs. Some studies have reported that a single-phase forms in HEAs when $0.80 \le \overline{Md} \le 2.60$ and that the range for eutectic HEAs is $0.95 \le \overline{Md} \le 1.20$ [94]. A critical $\overline{Md}$ value of 0.97 for FCC phase formation was determined in HEAs containing 3d transition elements only [95]. These studies indicated that a single solid solution phase can form when the $\overline{Md}$ value is within the critical value. A critical $\overline{Md}$ value of 0.992 for FCC-structured Co-free alloys including 4d transition elements is proposed in this work. The substitution of V, Mn and Mo a with a higher $\overline{Md}$ value for Co increases the critical value compared to other Co-containing alloys. While there is no assurance that Co-free alloys with $\overline{Md}$ values below a critical point are certainly a single FCC phase, the cooling rate may result in the formation of a minute quantity of the BCC phase [96]. Nevertheless, the $\overline{Md}$ method remains a dependable reference when formulating the composition of Co-free alloys.

### 4.3 Effects of *VEC* and $\overline{Md}$ on the mechanical properties of HEAs

The correlation between *VEC* and $\overline{Md}$ in relation to phase prediction can be readily observed in Figs. 3 and 4. To explore the effects of *VEC* and $\overline{Md}$ on the mechanical properties of HEAs, the hardness superimposed with *VEC* and $\overline{Md}$ are plotted in Fig. 7. The hardness of HEAs increases as *VEC* decreases and $\overline{Md}$ increases. Moreover, compared to the *VEC*, the correlation between $\overline{Md}$ and hardness displays greater convergence and closely aligns with the fitted line. This suggests a near-ideal positive correlation between the two variables. Considering this perspective, $\overline{Md}$ may serve as a superior reference compared to *VEC* when guiding the performance design of alloys. The *VEC* and $\overline{Md}$ methods have proven effective in the design of mechanical properties for other alloys as well as in predicting phase stability in some reports. It was found that the ductility of refractory HEAs composed of high melting point elements, such as Zr, Hf, V, Nb and Ta, can be improved by decreasing the *VEC* value [97]. Cast Ni alloys exhibited a maximum yield strength when $\overline{Md}$ and $\overline{Bo}$ (bond order) were 0.98 and 0.67, respectively [89]. The bond order and d-orbital energy level map for BCC phase stability evaluation of alloys confirms that the BCC phase undergoes transformation from a metastable to stable state as $\overline{Md}$ increases [98-100]. The stable BCC phase possesses higher strength and hardness and lower plasticity than the metastable BCC phase. Based on the experimental data in this work, the FCC structural HEAs demonstrate a similar trend of hardness increase with increasing $\overline{Md}$ values. This can be similarly explained by the effect of the d-orbital energy level on the stability of the FCC phase. The correlation between the mechanical properties and $\overline{Md}$ inspires us to design the composition in such a way that the $\overline{Md}$ value can be increased as much as possible to obtain a Co-free alloy with high strength while maintaining the single FCC phase.

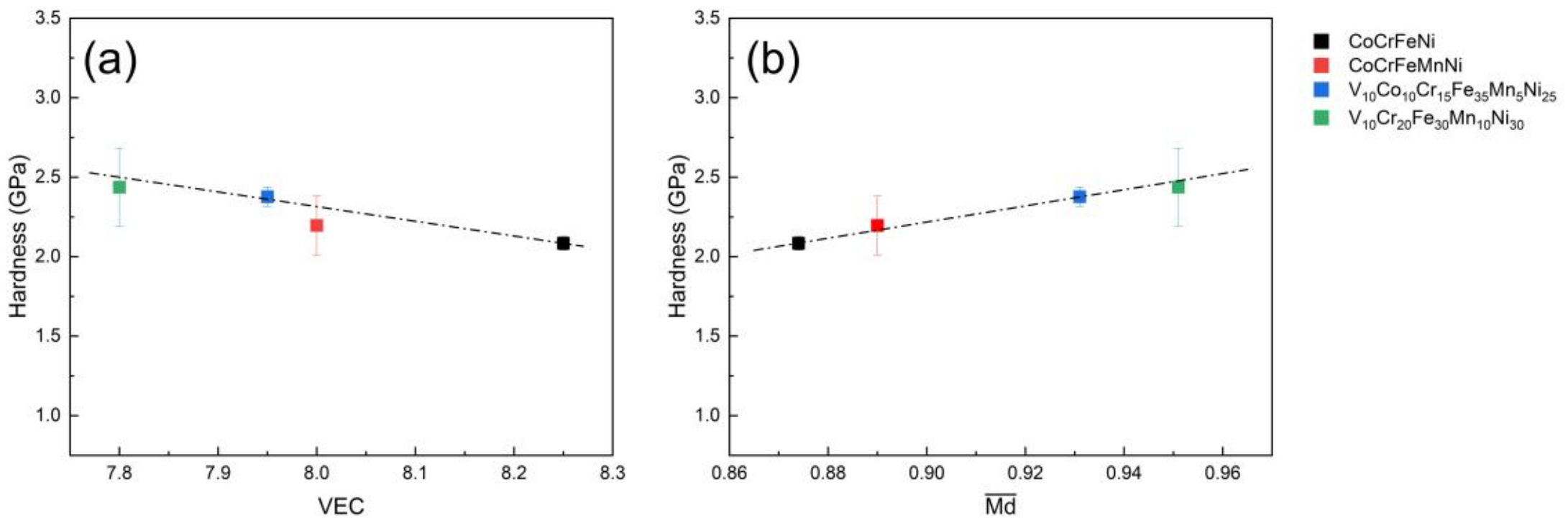

**Fig. 7** Relationship between hardness and (a) *VEC* and (b) $\overline{Md}$.

## 5. Conclusions

In this work, six parameters, $\overline{Md}$, *VEC*, $\Delta S_{mix}$, $\Delta H_{mix}$, $\Omega$ and $\delta$, were calculated to determine the solid solution phase, especially the FCC phase formation rules in Co-free HEAs. A new multicomponent system was designed to verify the formation rules. The conclusions are as follows:

(1) The solid solution phase in Co-free alloys forms when $0.787 \leq \overline{Md} \leq 1.260$, $6.46 \leq VEC \leq 9.00$, $9.59 \leq \Delta S_{mix} \leq 14.90$ J·K$^{-1}$·mol$^{-1}$, $-15.40 \leq \Delta H_{mix} \leq 7.66$ kJ·mol$^{-1}$, $\Omega \geq 1.54$ and $\delta \leq 7.08\%$ are satisfied simultaneously.

(2) The formation of the FCC phase is contingent upon satisfying specific criteria: $0.787 \leq \overline{Md} \leq 0.992$, $7.67 \leq VEC \leq 9.00$, $9.59 \leq \Delta S_{mix} \leq 13.38$ J·K$^{-1}$·mol$^{-1}$, $-11.00 \leq \Delta H_{mix} \leq 4.64$ kJ·mol$^{-1}$, $\delta \leq 5.30\%$, and $\Omega \geq 1.77$.

(3) Among the six parameters, the d-electron related $\overline{Md}$ and *VEC* are the critical factors that determine the FCC phase stability in Co-free alloys and $\overline{Md}$ can serve as a benchmark for developing mechanical properties.

(4) The newly designed alloy exhibits good mechanical properties with a hardness of 2.44 GPa and an elastic modulus of 162.88 GPa obtained by nanoindentation.

## Declaration of Competing Interest

The authors declare that they have no known competing financial interests or personal relationships that could have appeared to influence the work reported in this paper.

## Data availability

Data will be made available on request.

## CRediT author statement

Yulin Li: Investigation, Formal analysis, Visualization, Writing – original draft. Artur Olejarz: Investigation. Łukasz Kurpaska: Writing – review & editing. Eryang Lu: Formal analysis, Writing – review & editing. Mikko J. Alava: Writing – review & editing. Hyoung Seop Kim: Writing – review & editing. Wenyi Huo: Conceptualization, Methodology, Writing – original draft, Writing – review & editing, Supervision.

## Acknowledgments

This research is part of the project No. 2021/43/P/ST5/02663 co-funded by the National Science Centre and the European Union Framework Programme for Research and Innovation Horizon 2020 under the Marie Skłodowska-Curie grant agreement No. 945339. For the purpose of Open Access, the author has applied a CC-BY public copyright licence to any Author Accepted Manuscript (AAM) version arising from this submission. We also acknowledge support from the European Union Horizon 2020 research and innovation program under grant agreement no. 857470, the European Regional Development Fund via the Foundation for Polish Science International Research Agenda PLUS program grant No. MAB PLUS/2018/8, and the National Research Foundation of Korea (NRF) grant funded by the Korea government (MSIP) (NRF-2021R1A2C3006662, NRF-2022R1A5A1030054). The publication was created within the framework of the project of the Minister of Science and Higher Education "Support for the activities of Centres of Excellence established in Poland under Horizon 2020" under contract no. MEiN/2023/DIR/3795.

## References

[1] Y. Zhang, Reassembled nanoprecipitates resisting radiation, Nat. Mater. 22 (2023) 423-424.

[2] X. Wang, H. Huang, J. Shi, H. Xu, D. Meng, Recent progress of tungsten-based high-entropy alloys in nuclear fusion, Tungsten 3(2021) 143-160.

[3] Y. Li, Ł. Kurpaska, E. Lu, Z. Xie, H.S. Kim, W. Huo, Body-centered cubic phase stability in cobalt-free refractory high-entropy alloys, Res. Phys. 60 (2024) 107688.

[4] Y. Zhao, T. Yang, J. Zhu, D. Chen, Y. Yang, A. Hu, C. Liu, J.J. Kai, Development of high-strength Co-free high-entropy alloys hardened by nanosized precipitates, Scr. Mater. 148 (2018) 51-55.

[5] W. Huo, F. Fang, H. Zhou, Z. Xie, J. Shang, J. Jiang, Remarkable strength of CoCrFeNi high-entropy alloy wires at cryogenic and elevated temperatures, Scr. Mater. 141 (2017) 125-128.

[6] E. Ma, X. Wu, Tailoring heterogeneities in high-entropy alloys to promote strength-ductility synergy, Nat. Commun. 10(1) (2019) 5623.

[7] H. Du, J. Cai, Y. Wang, J. Yao, Q. Chen, Y. Cui, X. Liu, Effect of partial recrystallization on microstructure and tensile properties of NiFeCoCrMn high-entropy alloy, Trans. Nonferrous Met. Soc. China 32(3) (2022) 947-956.

[8] J. Pang, H. Zhang, Y. Ji, Z. Zhu, L. Zhang, H. Li, A. Wang, H. Zhang, High-temperature structural and mechanical stability of refractory high-entropy alloy $Nb_{40}Ti_{25}Al_{15}V_{10}Ta_5Hf_3W_2$, Mater. Charact. 205 (2023) 113321.

[9] W. Huo, H. Shi, X. Ren, J. Zhang, Microstructure and wear behavior of CoCrFeMnNbNi high-entropy alloy coating by TIG cladding, Adv. Mater. Sci. Eng. (2015) 647351.

[10] W. Huo, S. Wang, F. Fang, S. Tan, L. Kurpaska, Z. Xie, H.S. Kim, J. Jiang, Microstructure and corrosion resistance of highly <111> oriented electrodeposited CoNiFe medium-entropy alloy films, J. Mater. Res. Technol. 20 (2022) 1677-1684.

[11] Y. Zhang, G.M. Stocks, K. Jin, C. Lu, H. Bei, B.C. Sales, L. Wang, L.K. Béland, R.E. Stoller, G.D. Samolyuk, Influence of chemical disorder on energy dissipation and defect evolution in concentrated solid solution alloys, Nat. Commun. 6(1) (2015) 8736.

[12] J. Pang, B. Wei, H. Zhang, Y. Ji, Z. Zhu, L. Zhang, H. Fu, H. Li, A. Wang, H. Zhang, Improvement in corrosion resistance of biocompatible $Ti_{1.5}Al_{0.3}ZrNb$ refractory high entropy alloy in simulated body fluid by nanosecond laser shock processing, Corros. Sci. 224 (2023) 111484.

[13] W. Zhu, W. Huo, S. Wang, X. Wang, K. Ren, S. Tan, F. Fang, Z. Xie, J. Jiang, Phase formation prediction of high-entropy alloys: a deep learning study, J. Mater. Res. Technol. 18 (2022) 800-809.

[14] W. Zhu, W. Huo, S. Wang, L. Kurpaska, F. Fang, S. Papanikolaou, H.S. Kim, J. Jiang, Machine learning-based hardness prediction of high-entropy alloys for laser additive manufacturing. JOM 75 (2023) 5537-5548.

[15] N.A.P.K. Kumar, C. Li, K.J. Leonard, H. Bei, S.J. Zinkle, Microstructural stability and mechanical behavior of FeNiMnCr high entropy alloy under ion irradiation, Acta Mater. 113 (2016) 230-244.
[16] C. Lu, T. Yang, K. Jin, N. Gao, P. Xiu, Y. Zhang, F. Gao, H. Bei, W.J. Weber, K. Sun, Radiation-induced segregation on defect clusters in single-phase concentrated solid-solution alloys, Acta Mater. 127 (2017) 98-107.
[17] Y. Chen, D. Chen, J. Weaver, J. Gigax, Y. Wang, N.A. Mara, S. Fensin, S.A. Maloy, A. Misra, N. Li, Heavy ion irradiation effects on CrFeMnNi and AlCrFeMnNi high entropy alloys, J. Nucl. Mater. 574 (2023) 154163.
[18] C. Li, X. Hu, T. Yang, N.K. Kumar, B.D. Wirth, S.J. Zinkle, Neutron irradiation response of a Co-free high entropy alloy, J. Nucl. Mater. 527 (2019) 151838.
[19] Y. Zhang, Y.J. Zhou, J.P. Lin, G.L. Chen, P.K. Liaw, Solid-solution phase formation rules for multi-component alloys, Adv. Eng. Mater. 10(6) (2008) 534-538.
[20] S. Guo, C. Ng, J. Lu, C. Liu, Effect of valence electron concentration on stability of fcc or bcc phase in high entropy alloys, J. Appl. Phys. 109(10) (2011) 103505.
[21] Z. Xing, J. Pang, H. Zhang, Y. Ji, Z. Zhu, A. Wang, L. Zhang, H. Li, H. Fu, H. Zhang, Optimizing the microstructure and mechanical performance of Fe-Ni-Cr-Al high entropy alloys via Ti addition, J. Alloys Compd. 943 (2023) 169149.
[22] Y. Lu, Y. Dong, L. Jiang, T. Wang, T. Li, Y. Zhang, A criterion for topological close-packed phase formation in high entropy alloys, Entropy 17(4) (2015) 2355-2366.
[23] C.J. Tong, Y.L. Chen, J.W. Yeh, S.J. Lin, S.K. Chen, T.T. Shun, C.H. Tsau, S.Y. Chang, Microstructure characterization of $Al_x$CoCrCuFeNi high-entropy alloy system with multiprincipal elements, Metall. Mater. Trans. A 36 (2005) 881-893.
[24] W. Du, N. Liu, Z. Peng, P. Zhou, H. Xiang, X. Wang, The effect of Ti addition on phase selection of $CoCrCu_{0.5}FeNi$ high-entropy alloys, Mater. Sci. Technol. 34(4) (2018) 473-479.
[25] Y. Chou, J. Yeh, H. Shih, The effect of molybdenum on the corrosion behaviour of the high-entropy alloys $Co_{1.5}CrFeNi_{1.5}Ti_{0.5}Mo_x$ in aqueous environments, Corros. Sci. 52(8) (2010) 2571-2581.
[26] Y. Zhou, Y. Zhang, Y. Wang, G. Chen, Solid solution alloys of $AlCoCrFeNiTi_x$ with excellent room-temperature mechanical properties, Appl. Phys. Lett. 90(18) (2007) 181904.
[27] C.C. Tung, J.W. Yeh, T.T. Shun, S.K. Chen, Y.S. Huang, H.C. Chen, On the elemental effect of AlCoCrCuFeNi high-entropy alloy system, Mater. Lett. 61(1) (2007) 1-5.
[28] X. Wang, Y. Zhang, Y. Qiao, G. Chen, Novel microstructure and properties of multicomponent $CoCrCuFeNiTi_x$ alloys, Intermetallics 15(3) (2007) 357-362.
[29] J. Zhu, H. Zhang, H. Fu, A. Wang, H. Li, Z. Hu, Microstructures and compressive properties of multicomponent $AlCoCrCuFeNiMo_x$ alloys, J. Alloys Compd. 497(1-2) (2010) 52-56.

[30] M.S. Mehranpour, H. Shahmir, H.S. Kim, Microstructure tailoring by manipulating chemical composition in novel CoNiMnCrAl high-entropy alloys, J. Alloys Compd. 944 (2023) 169207.
[31] Z. Yunjun, Z. Yong, W. Yanli, C. Guoliang, Microstructure characterization of $Al_{x}(TiVCrMnFeCoNiCu)_{(100-x)}$ high-entropy alloy system with multi-principal elements, Rare Met. Mater. Eng. 36(12) (2007) 2136-2139.
[32] M.R. Chen, S.J. Lin, J.W. Yeh, S.K. Chen, Y.S. Huang, C.P. Tu, Microstructure and properties of $Al_{0.5}CoCrCuFeNiTi_{x}$($x$=0-2.0) high-entropy alloys, Mater. Trans. 47(5) (2006) 1395-1401.
[33] C.Y. Hsu, W.R. Wang, W.Y. Tang, S.K. Chen, J.W. Yeh, Microstructure and mechanical properties of new $AlCo_{x}CrFeMo_{0.5}Ni$ High-Entropy Alloys, Adv. Eng. Mater. 12(1-2) (2010) 44-49.
[34] G.Y. Ke, S.K. Chen, T. Hsu, J.W. Yeh, FCC and BCC equivalents in as-cast solid solutions of $Al_{x}Co_{y}Cr_{z}Cu_{0.5}Fe_{v}Ni_{w}$ high-entropy alloys, Ann. Chim., 2006, 669-683.
[35] M.R. Chen, S.J. Lin, J.W. Yeh, M.H. Chuang, S.K. Chen, Y.S. Huang, Effect of vanadium addition on the microstructure, hardness, and wear resistance of $Al_{0.5}CoCrCuFeNi$ high-entropy alloy, Metall. Mater. Trans. A 37 (2006) 1363-1369.
[36] C.Y. Hsu, T.S. Sheu, J.W. Yeh, S.K. Chen, Effect of iron content on wear behavior of $AlCoCrFe_{x}Mo_{0.5}Ni$ high-entropy alloys, Wear 268(5-6) (2010) 653-659.
[37] C. Li, J. Li, M. Zhao, Q. Jiang, Effect of alloying elements on microstructure and properties of multiprincipal elements high-entropy alloys, J. Alloys Compd. 475(1-2) (2009) 752-757.
[38] M. Zhang, X. Shi, Z. Li, H. Xu, G. Li, Corrosion behaviors and mechanism of $CrFeNi_{2}$ based high-entropy alloys, Corros. Sci. 207 (2022) 110562.
[39] T. Borkar, V. Chaudhary, B. Gwalani, D. Choudhuri, C.V. Mikler, V. Soni, T. Alam, R. V. Ramanujan, R. Banerjee, A combinatorial approach for assessing the magnetic properties of high entropy alloys: role of Cr in $AlCo_{x}Cr_{1-x}FeNi$, Adv. Eng. Mater. 19(8) (2017) 1700048.
[40] B. Ren, Z. Liu, B. Cai, M. Wang, L. Shi, Aging behavior of a $CuCr_{2}Fe_{2}NiMn$ high-entropy alloy, Mater. Des. 33 (2012) 121-126.
[41] S. Guo, C. Ng, C.T. Liu, Anomalous solidification microstructures in Co-free $Al_{x}CrCuFeNi_{2}$ high-entropy alloys, J. Alloys Compd. 557 (2013) 77-81.
[42] X. Huang, J. Miao, S. Li, C.D. Taylor, A.A. Luo, Co-free CuFeMnNi high-entropy alloy with tunable tensile properties by thermomechanical processing, J. Mater. Sci. 56 (2021) 7670-7680.
[43] B. Ren, Z. Liu, D. Li, L. Shi, B. Cai, M. Wang, Corrosion behavior of CuCrFeNiMn high entropy alloy system in 1 M sulfuric acid solution, Mater. Corros. 63(9) (2012) 828-834.
[44] M.H. Tsai, K.Y. Tsai, C.W. Tsai, C. Lee, C.C. Juan, J.W. Yeh, Criterion for sigma phase formation in Cr- and V-containing high-entropy alloys, Mater. Res. Lett. 1(4) (2013) 207-212.

[45] N. Stepanov, D. Shaysultanov, R. Chernichenko, M. Tikhonovsky, S. Zherebtsov, Effect of Al on structure and mechanical properties of Fe-Mn-Cr-Ni-Al non-equiatomic high entropy alloys with high Fe content, J. Alloys Compd. 770 (2019) 194-203.
[46] L. Bai, Y. Wang, Y. Yan, X. Li, Y. Lv, J. Chen, Effect of carbon on microstructure and mechanical properties of $Fe_{36}Mn_{36}Ni_{9}Cr_{9}Al_{10}$ high-entropy alloys, Mater. Sci. Technol. 36(17) (2020) 1851-1860.
[47] V. Voyevodin, S. Karpov, G. Tolstolutskaya, M. Tikhonovsky, A. Velikodnyi, I. Kopanets, G. Tolmachova, A. Kalchenko, R. Vasilenko, I. Kolodiy, Effect of irradiation on microstructure and hardening of Cr-Fe-Ni-Mn high-entropy alloy and its strengthened version, Philos. Mag. 100(7) (2020) 822-836.
[48] H. Chen, C.W. Tsai, C.C. Tung, L. Yeh, T.T. Shun, C.C. Yang, S.K. Chen, Effect of the substitution of Co by Mn in Al-Cr-Cu-Fe-Co-Ni high entropy alloys, Ann. Chim., 2006, 685-698.
[49] L. Bai, Y. Liu, Y. Guo, Y. Lv, T. Guo, J. Chen, Effects of Al addition on microstructure and mechanical properties of Co-free $(Fe_{40}Mn_{40}Ni_{10}Cr_{10})_{100-x}Al_{x}$ high-entropy alloys, J. Alloys Compd. 879 (2021) 160342.
[50] H. Zhang, C. Li, Z. Zhu, H. Huang, Y. Lu, T. Wang, T. Li, Effects of He-ion irradiation on the microstructures and mechanical properties of the novel Co-free $V_{x}CrFeMnNi_{y}$ high-entropy alloys, J. Nucl. Mater. 572 (2022) 154074.
[51] A. Munitz, L. Meshi, M. Kaufman, Heat treatments' effects on the microstructure and mechanical properties of an equiatomic Al-Cr-Fe-Mn-Ni high entropy alloy, Mater. Sci. Eng. A 689 (2017) 384-394.
[52] M. Wu, I. Baker, High strength and high ductility in a novel $Fe_{40.2}Ni_{11.3}Mn_{30}Al_{7.5}Cr_{11}$ multiphase high entropy alloy, J. Alloys Compd. 820 (2020) 153181.
[53] D. Vogiatzief, A. Evirgen, M. Pedersen, U. Hecht, Laser powder bed fusion of an Al-Cr-Fe-Ni high-entropy alloy produced by blending of prealloyed and elemental powder: Process parameters, microstructures and mechanical properties, J. Alloys Compd. 918 (2022) 165658.
[54] C.W. Lin, M.H. Tsai, C.W. Tsai, J.W. Yeh, S.K. Chen, Microstructure and aging behaviour of $Al_{5}Cr_{32}Fe_{35}Ni_{22}Ti_{6}$ high entropy alloy, Mater. Sci. Technol. 31(10) (2015) 1165-1170.
[55] M. Chen, Y. Liu, Y. Li, X. Chen, Microstructure and mechanical properties of $AlTiFeNiCuCr_{x}$ high-entropy alloy with multi-principal elements, Acta Metall. Sin. 43(10) (2007) 1020-1024.
[56] J. Zeng, C. Wu, H. Peng, Y. Liu, J. Wang, X. Su, Microstructure and microhardness of as-cast and 800 °C annealed $Al_{x}Cr_{0.2}Fe_{0.2}Ni_{0.6-x}$ and $Al_{0.2}Cr_{0.2}Fe_{y}Ni_{0.6-y}$ alloys, Vacuum 152 (2018) 214-221.
[57] R.K. Nutor, M. Azeemullah, Q. Cao, X. Wang, D. Zhang, J. Jiang, Microstructure and properties of a Co-free $Fe_{50}Mn_{27}Ni_{10}Cr_{13}$ high entropy alloy, J. Alloys Compd. 851 (2021) 156842.
[58] A.K. Singh, A. Subramaniam, On the formation of disordered solid solutions in multi-component alloys, J. Alloys Compd. 587 (2014) 113-119.

[59] A. Durga, K. Hari Kumar, B. Murty, Phase formation in equiatomic high entropy alloys: CALPHAD approach and experimental studies, Trans. Indian Inst. Met. 65 (2012) 375-380.
[60] L. Wang, L. Wang, S. Zhou, Q. Xiao, Y. Xiao, X. Wang, T. Cao, Y. Ren, Y.J. Liang, L. Wang, Precipitation and micromechanical behavior of the coherent ordered nanoprecipitation strengthened Al-Cr-Fe-Ni-V high entropy alloy, Acta Mater. 216 (2021) 117121.
[61] F. Otto, Y. Yang, H. Bei, E.P. George, Relative effects of enthalpy and entropy on the phase stability of equiatomic high-entropy alloys, Acta Mater. 61(7) (2013) 2628-2638.
[62] K.C. Hsieh, C.F. Yu, W.T. Hsieh, W.R. Chiang, J.S. Ku, J.H. Lai, C.P. Tu, C.C. Yang, The microstructure and phase equilibrium of new high performance high-entropy alloys, J. Alloys Compd. 483(1-2) (2009) 209-212.
[63] Y. Wu, X. Jin, M. Zhang, H. Yang, J. Qiao, Y. Wu, Yield strength-ductility trade-off breakthrough in Co-free $Fe_{40}Mn_{10}Cr_{25}Ni_{25}$ high-entropy alloys with partial recrystallization, Mater. Today Commun. 28 (2021) 102718.
[64] L.H. Wen, H.C. Kou, J.S. Li, H. Chang, X.Y. Xue, L. Zhou, Effect of aging temperature on microstructure and properties of AlCoCrCuFeNi high-entropy alloy, Intermetallics 17(4) (2009) 266-269.
[65] Y. Yang, J. Pang, H. Zhang, A. Wang, Z. Zhu, H. Li, G. Tang, L. Zhang, H. Zhang, Short-Term Splitting and Long-Term Stability of Cuboidal Nanoparticles in $Ni_{44}Co_{22}Cr_{22}Al_6Nb_6$ Multi-Principal Element Alloy, Acta Metall. Sin. 36(6) (2023) 999-1006.
[66] Y. Yang, J. Pang, Z. Zhang, Y. Wang, Y. Ji, Z. Zhu, L. Zhang, A. Wang, H. Zhang, H. Zhang, Strength-ductility synergy and superior strain-hardening ability of $Ni_{38}Co_{25}Fe_{13}Cr_{10}Al_7Ti_7$ multi principal element alloy through heterogeneous L12 structure modulation, J. Alloys Compd. 984 (2024) 173916.
[67] M. Morinaga, N. Yukawa, H. Adachi, H. Ezaki, New PHACOMP and its applications to alloy design, Superalloys 1984 (1984) 523-532.
[68] T.B. Massalski, Comments concerning some features of phase diagrams and phase transformations, Mater. Trans. 51(4) (2010) 583-596.
[69] G. Sheng, C.T. Liu, Phase stability in high entropy alloys: Formation of solid-solution phase or amorphous phase, Prog. Nat. Sci.: Mater. Int. 21(6) (2011) 433-446.
[70] I. Toda-Caraballo, P. Rivera-Díaz-del-Castillo, A criterion for the formation of high entropy alloys based on lattice distortion, Intermetallics 71 (2016) 76-87.
[71] X. Yang, Y. Zhang, Prediction of high-entropy stabilized solid-solution in multi-component alloys, Mater. Chem. Phys. 132(2-3) (2012) 233-238.
[72] A. Takeuchi, A. Inoue, Classification of bulk metallic glasses by atomic size difference, heat of mixing and period of constituent elements and its application to characterization of the main alloying element, Mater. Trans. 46(12) (2005) 2817-2829.
[73] A. Gali, E.P. George, Tensile properties of high- and medium-entropy alloys, Intermetallics 39 (2013) 74-78.

[74] B. Cantor, I.T.H. Chang, P. Knight, A.J.B. Vincent, Microstructural development in equiatomic multicomponent alloys, Mater. Sci. Eng. A 375-377 (2004) 213-218.

[75] Y.H. Jo, S. Jung, W.M. Choi, S.S. Sohn, H.S. Kim, B.J. Lee, N.J. Kim, S. Lee, Cryogenic strength improvement by utilizing room-temperature deformation twinning in a partially recrystallized VCrMnFeCoNi high-entropy alloy, Nat. Commun. 8(1) (2017) 15719.

[76] R.W. Cahn, P. Haasen, Phys. Metall., Elsevier 1996.

[77] O.R. Deluigi, R.C. Pasianot, F. Valencia, A. Caro, D. Farkas, E.M. Bringa, Simulations of primary damage in a High Entropy Alloy: Probing enhanced radiation resistance, Acta Mater. 213 (2021) 116951.

[78] Z. Li, S. Zhao, R.O. Ritchie, M.A. Meyers, Mechanical properties of high-entropy alloys with emphasis on face-centered cubic alloys, Prog. Mater. Sci. 102 (2019) 296-345.

[79] J. Ding, M. Asta, R.O. Ritchie, Melts of CrCoNi-based high-entropy alloys: Atomic diffusion and electronic/atomic structure from ab initio simulation, Appl. Phys. Lett. 113(11) (2018) 111902.

[80] N.K. Kumar, C. Li, K. Leonard, H. Bei, S. Zinkle, Microstructural stability and mechanical behavior of FeNiMnCr high entropy alloy under ion irradiation, Acta Mater. 113 (2016) 230-244.

[81] S. Guo, Q. Hu, C. Ng, C. Liu, More than entropy in high-entropy alloys: Forming solid solutions or amorphous phase, Intermetallics 41 (2013) 96-103.

[82] Y. Zhang, X. Yang, P. Liaw, Alloy design and properties optimization of high-entropy alloys, JOM 64 (2012) 830-838.

[83] M.H. Tsai, J.H. Li, A.C. Fan, P.H. Tsai, Incorrect predictions of simple solid solution high entropy alloys: Cause and possible solution, Scr. Mater. 127 (2017) 6-9.

[84] O. Senkov, D. Miracle, A new thermodynamic parameter to predict formation of solid solution or intermetallic phases in high entropy alloys, J. Alloys Compd. 658 (2016) 603-607.

[85] S. Guo, Phase selection rules for cast high entropy alloys: an overview, Mater. Sci. Technol. 31(10) (2015) 1223-1230.

[86] M.G. Poletti, L. Battezzati, Electronic and thermodynamic criteria for the occurrence of high entropy alloys in metallic systems, Acta Mater. 75 (2014) 297-306.

[87] U. Mizutani, Hume-Rothery rules for structurally complex alloy phases. MRS Bulletin 37 (2012) 169.

[88] M. Morinaga, N. Yukawa, H. Ezaki, H. Adachi, Solid solubilities in transition-metal-based fcc alloys, Philos. Mag. A 51(2) (1985) 223-246.

[89] H. Adachi, T. Mukoyama, J. Kawai, Hartree-Fock-Slater method for materials science: the DV-X Alpha method for design and characterization of materials, Springer Science & Business Media 2006.

[90] M. Morinaga, Alloy design based on molecular orbital method, Mater. Trans. 57(3) (2016) 213-226.

[91] Y. Matsumoto, M. Morinaga, T. Nambu, T. Sakaki, Alloying effects on the electronic structure of chromium, J. Phys.: Condens.Matter 8(20) (1996) 3619.
[92] S.S. Sidhu, H. Singh, M.A.H. Gepreel, A review on alloy design, biological response, and strengthening of β-titanium alloys as biomaterials, Mater. Sci. Eng. C 121 (2021) 111661.
[93] M. Morinaga, H. Yukawa, Alloy design with the aid of molecular orbital method, Bull. Mater. Sci. 20 (1997) 805-815.
[94] T. Li, Y. Lu, T. Wang, T. Li, Grouping strategy via d-orbit energy level to design eutectic high-entropy alloys, Appl. Phys. Lett. 119(7) (2021) 071905.
[95] S. Sheikh, U. Klement, S. Guo, Predicting the solid solubility limit in high-entropy alloys using the molecular orbital approach, J. Appl. Phys. 118(19) (2015).
[96] S. Singh, N. Wanderka, B. Murty, U. Glatzel, J. Banhart, Decomposition in multi-component AlCoCrCuFeNi high-entropy alloy, Acta Mater. 59(1) (2011) 182-190.
[97] S. Sheikh, S. Shafeie, Q. Hu, J. Ahlström, C. Persson, J. Veselý, J. Zýka, U. Klement, S. Guo, Alloy design for intrinsically ductile refractory high-entropy alloys, J. Appl. Phys. 120(16) (2016) 194902.
[98] X. Wen, Y. Wu, H. Huang, S. Jiang, H. Wang, X. Liu, Y. Zhang, X. Wang, Z. Lu, Effects of Nb on deformation-induced transformation and mechanical properties of $HfNb_xTa_{0.2}TiZr$ high entropy alloys, Mater. Sci. Eng. A 805 (2021) 140798.
[99] L. Lilensten, J.-P. Couzinié, J. Bourgon, L. Perrière, G. Dirras, F. Prima, I. Guillot, Design and tensile properties of a bcc Ti-rich high-entropy alloy with transformation-induced plasticity, Mater. Res. Lett. 5(2) (2017) 110-116.
[100] Y. Huang, J. Gao, S. Wang, D. Guan, Y. Xu, X. Hu, W.M. Rainforth, Q. Zhu, I. Todd, Influence of tantalum composition on mechanical behavior and deformation mechanisms of $TiZrHfTa_x$ high entropy alloys, J. Alloys Compd. 903 (2022) 163796.

# Appendix

**Table. A1** Compositions of Co-containing multi-component alloys and parameters $\overline{Md}$, *VEC*, $\Delta S_{mix}$, $\Delta H_{mix}$, $\Omega$, $\delta$ and phases. Notation: the intermetallic phases are denoted as IM.

| Materials | $\overline{Md}$ | *VEC* | $\Delta Smix$ (J·K-1·mol-1) | $\Delta Hmix$ (kJ·mol-1) | $\Omega$ | $\delta$ | Phase | Ref. |
|---|---|---|---|---|---|---|---|---|
| $Al_{0.3}CoCrNiCuFe$ | 0.883 | 8.47 | 14.43 | 0.16 | 158.62 | 3.44 | FCC | [23] |
| $Al_{0.5}CoCrCu_{0.5}FeNi$ | 0.950 | 8.00 | 14.53 | -4.60 | 5.45 | 4.37 | FCC | [34] |
| $Al_{0.5}CoCrCuFeNi$ | 0.920 | 8.27 | 14.70 | -1.52 | 16.36 | 4.17 | FCC | [27] |
| $Al_{0.5}CoCrCuFeNiTi_{0.8}$ | 1.091 | 7.73 | 16.00 | -10.11 | 2.73 | 6.24 | BCC+FCC+IM | [32] |
| $Al_{0.5}CoCrCuFeNiTi_{1.0}$ | 1.128 | 7.62 | 16.01 | -11.60 | 2.39 | 6.52 | BCC+FCC+IM | [32] |
| $Al_{0.5}CoCrCuFeNiTi_{1.2}$ | 1.162 | 7.51 | 15.97 | -12.89 | 2.15 | 6.73 | BCC+FCC+IM | [32] |
| $Al_{0.5}CoCrCuFeNiV$ | 1.016 | 7.77 | 16.01 | -5.25 | 5.39 | 4.04 | BCC+FCC | [35] |
| $Al_{0.5}CoCrCuFeNiV_{0.2}$ | 0.942 | 8.16 | 15.45 | -2.50 | 10.57 | 4.14 | FCC | [35] |
| $Al_{0.5}CoCrCuFeNiV_{0.4}$ | 0.962 | 8.05 | 15.76 | -3.34 | 8.14 | 4.11 | BCC+FCC | [35] |
| $Al_{0.5}CoCrCuFeNiV_{0.6}$ | 0.981 | 7.95 | 15.92 | -4.07 | 6.81 | 4.09 | BCC+FCC+IM | [35] |
| $Al_{0.5}CoCrCuFeNiV_{0.8}$ | 0.999 | 7.86 | 16.00 | -4.71 | 5.97 | 4.06 | BCC+FCC+IM | [35] |
| $Al_{0.5}CrFeCoNi$ | 0.988 | 7.67 | 13.15 | -9.09 | 2.56 | 4.59 | FCC | [19] |
| $Al_{0.8}FeCoNiCrCu$ | 0.971 | 8.00 | 14.87 | -3.61 | 6.80 | 4.92 | BCC+FCC | [23] |
| $Al_{1.5}CoCrCu_{0.5}FeNi$ | 1.109 | 7.17 | 14.53 | -10.14 | 2.29 | 6.11 | BCC+FCC | [34] |
| $Al_{2.0}FeCoNiCrCu$ | 1.130 | 7.14 | 14.53 | -8.65 | 2.57 | 6.24 | BCC+FCC | [23] |
| $Al_{2.3}FeCoNiCrCu$ | 1.162 | 6.97 | 14.35 | -9.38 | 2.30 | 6.39 | BCC+FCC | [23] |
| $Al_{2.8}FeCoNiCrCu$ | 1.209 | 6.72 | 14.01 | -10.28 | 2.00 | 6.55 | BCC | [23] |
| $Al_2CoCrCu_{0.5}FeNi$ | 1.169 | 6.85 | 14.23 | -11.60 | 1.89 | 6.45 | BCC+FCC | [34] |
| $Al_{3.0}FeCoNiCrCu$ | 1.226 | 6.63 | 13.86 | -10.56 | 1.91 | 6.60 | BCC | [23] |
| $Al_5Cr_{10}Co_{25}Mn_{30}Ni_{30}$ | 0.906 | 8.10 | 12.05 | -8.99 | 2.25 | 4.40 | FCC | [30] |
| $Al_5Cr_{10}Co_{30}Mn_{25}Ni_{30}$ | 0.897 | 8.20 | 12.05 | -8.63 | 2.36 | 4.31 | FCC | [30] |
| $Al_5Cr_{10}Co_{30}Mn_{30}Ni_{25}$ | 0.909 | 8.05 | 12.05 | -8.72 | 2.32 | 4.38 | FCC | [30] |
| $AlCo_{0.5}CrCu_{0.5}FeNi$ | 1.063 | 7.40 | 14.53 | -7.92 | 3.02 | 5.70 | BCC+FCC | [34] |
| $AlCo_{0.5}CrCuFeNi$ | 1.022 | 7.73 | 14.70 | -4.50 | 5.29 | 5.44 | BCC+FCC | [34] |
| $AlCo_{1.5}CrCu_{0.5}FeNi$ | 1.015 | 7.67 | 14.53 | -7.83 | 3.09 | 5.32 | BCC+FCC | [34] |
| $AlCo_2CrCu_{0.5}FeNi$ | 0.997 | 7.77 | 14.23 | -7.67 | 3.10 | 5.17 | BCC+FCC | [34] |
| $AlCo_{3.5}CrCu_{0.5}FeNi$ | 0.956 | 8.00 | 13.09 | -7.03 | 3.15 | 4.74 | BCC+FCC | [34] |
| $AlCo_3CrCu_{0.5}FeNi$ | 0.967 | 7.93 | 13.48 | -7.25 | 3.13 | 4.87 | BCC+FCC | [34] |
| $AlCoCr_{0.5}Cu_{0.5}FeNi$ | 1.026 | 7.70 | 14.53 | -8.32 | 2.80 | 5.34 | BCC+FCC | [34] |
| $AlCoCr_{0.5}CuFeNi$ | 0.989 | 8.00 | 14.70 | -5.02 | 4.62 | 5.43 | BCC+FCC | [27] |
| $AlCoCr_{1.5}Cu_{0.5}FeNi$ | 1.045 | 7.42 | 14.53 | -7.56 | 3.27 | 5.34 | BCC+FCC | [34] |
| $AlCoCrCu_{0.5}FeNi$ | 1.037 | 7.55 | 14.70 | -7.93 | 3.06 | 5.50 | BCC | [27] |
| $AlCoCrCuFe_{0.5}Ni$ | 1.015 | 7.82 | 14.70 | -5.55 | 4.27 | 5.39 | BCC+FCC | [27] |
| AlCoCrCuFeNi | 1.002 | 7.83 | 14.90 | -4.78 | 5.08 | 5.28 | BCC+FCC | [34] |
| $AlCoCrCuFeNi_{0.5}$ | 1.027 | 7.64 | 14.70 | -3.90 | 6.11 | 5.42 | BCC+FCC | [27] |
| AlCoCrFeNi | 1.079 | 7.20 | 13.38 | -12.32 | 1.83 | 5.76 | FCC+B2 | [37] |
| $AlCoCu_{0.5}FeNi$ | 1.013 | 7.89 | 13.15 | -8.69 | 2.33 | 5.89 | BCC+FCC | [34] |
| $AlCrFeCoNiCu_{0.25}$ | 1.057 | 7.38 | 14.34 | -9.94 | 2.41 | 5.63 | BCC | [26] |
| $Co_{0.5}CrFeNiAlMo_{0.5}$ | 1.156 | 6.90 | 14.53 | -11.72 | 2.23 | 5.91 | BCC+IM | [33] |
| $Co_{1.5}CrFeNiAlMo_{0.5}$ | 1.093 | 7.25 | 14.53 | -11.08 | 2.35 | 5.58 | BCC+IM | [33] |
| $Co_2CrFeNiAlMo_{0.5}$ | 1.069 | 7.38 | 14.23 | -10.70 | 2.38 | 5.42 | BCC+FCC+IM | [33] |
| $CoCrCu_{0.5}FeNi$ | 0.845 | 8.56 | 13.15 | 0.49 | 48.30 | 0.77 | FCC | [24] |
| $CoCrCu_{0.5}FeNiTi$ | 1.104 | 7.73 | 14.70 | -11.77 | 2.29 | 6.37 | FCC+IM | [24] |
| $CoCrCu_{0.5}FeNiTi_{0.1}$ | 0.876 | 8.46 | 13.73 | -1.26 | 19.85 | 2.59 | FCC | [24] |
| $CoCrCu_{0.5}FeNiTi_{0.3}$ | 0.934 | 8.27 | 14.27 | -4.33 | 6.00 | 4.11 | FCC | [24] |
| $CoCrCu_{0.5}FeNiTi_{0.5}$ | 0.987 | 8.10 | 14.53 | -6.92 | 3.84 | 5.04 | FCC+B2+IM | [24] |
| CoCrCuFeMnNiTiV | 1.110 | 7.50 | 17.29 | -8.13 | 3.85 | 5.49 | BCC+FCC+IM | [31] |
| CoCrCuFeNiTi | 1.063 | 8.00 | 14.90 | -8.44 | 3.17 | 6.11 | FCC+IM | [28] |
| $CoCrCuFeNiTi_{0.1}$ | 0.850 | 8.71 | 13.92 | 1.60 | 15.43 | 2.51 | FCC | [28] |
| $CoCrCuFeNiTi_{0.5}$ | 0.954 | 8.36 | 14.70 | -3.70 | 7.08 | 4.82 | FCC | [28] |
| $CoCrCuFeNiTi_{0.8}$ | 1.022 | 8.14 | 14.87 | -6.75 | 3.95 | 5.68 | FCC+IM | [28] |
| $CoCrFe_{0.6}NiAlMo_{0.5}$ | 1.142 | 7.02 | 14.61 | -12.32 | 2.13 | 5.83 | BCC+IM | [36] |
| $CoCrFe_{1.5}NiAlMo_{0.5}$ | 1.100 | 7.17 | 14.53 | -10.50 | 2.49 | 5.61 | BCC+IM | [36] |
| $CoCrFe_2NiAlMo_{0.5}$ | 1.081 | 7.23 | 14.23 | -9.70 | 2.64 | 5.50 | BCC+IM | [36] |
| $CoCrFeNi_2$ | 0.842 | 8.60 | 11.08 | -3.84 | 5.32 | 0.00 | FCC | [38] |
| $CoCrFeNiAlMo_{0.5}$ | 1.122 | 7.09 | 14.70 | -11.44 | 2.31 | 5.74 | BCC+IM | [36] |
| CoCrFeNiCuAlMo | 1.080 | 7.57 | 16.18 | -3.51 | 8.34 | 5.29 | BCC+IM | [29] |
| $CoCrFeNiCuAlMo_{0.4}$ | 1.036 | 7.72 | 15.91 | -4.20 | 6.47 | 5.30 | BCC+IM | [29] |
| $CoCrFeNiCuAlMo_{0.6}$ | 1.051 | 7.67 | 16.08 | -3.95 | 7.10 | 5.31 | BCC+IM | [29] |
| $CoCrFeNiCuAlMo_{0.8}$ | 1.066 | 7.62 | 16.16 | -3.72 | 7.73 | 5.30 | BCC+IM | [29] |
| FeCoNiCrCu | 0.822 | 8.80 | 13.38 | 3.20 | 7.40 | 1.05 | FCC | [23] |
| $FeCoNiCrCuAl_{1.5}$ | 1.071 | 7.46 | 14.78 | -7.05 | 3.30 | 5.88 | BCC+FCC | [23] |
| $Ti_{0.5}Co_{1.5}CrFeNi_{1.5}Mo_{0.5}$ | 1.025 | 7.92 | 14.17 | -10.25 | 2.69 | 5.09 | FCC+IM | [25] |
| $Ti_{0.5}Co_{1.5}CrFeNi_{1.5}Mo_{0.8}$ | 1.050 | 7.83 | 14.39 | -9.96 | 2.87 | 5.17 | FCC+IM | [25] |
| VCuFeCoNi | 0.902 | 8.60 | 13.38 | -2.24 | 10.57 | 2.21 | FCC | [26] |

**Table. A2** Compositions of Co-free multi-component alloys and parameters $\overline{Md}$, $VEC$, $\Delta S_{mix}$, $\Delta H_{mix}$, $\Omega$, $\delta$ and phases. Notation: the intermetallic phases are denoted as IM.

| Materials | $\overline{Md}$ | $VEC$ | $\Delta Smix$ (J·K-1·mol-1) | $\Delta Hmix$ (kJ·mol-1) | $\Omega$ | $\delta$ | Phase | Ref. |
|---|---|---|---|---|---|---|---|---|
| $Al_{0.2}CrCuFe$ | 0.936 | 8.00 | 10.51 | 7.66 | 2.37 | 3.55 | BCC+FCC | [20] |
| $Al_{0.2}CrCuFeNi_2$ | 0.852 | 8.77 | 12.01 | 0.12 | 175.41 | 2.92 | FCC | [20] |
| $Al_{0.3}CrCuFeMnNi$ | 0.917 | 8.09 | 14.43 | -0.27 | 89.33 | 4.20 | BCC+FCC | [48] |
| $Al_{0.3}CrFe_{1.5}MnNi_{0.5}$ | 1.003 | 7.19 | 12.32 | -5.51 | 3.93 | 4.69 | BCC+FCC+IM | [62] |
| $Al_{0.4}CrCuFeNi_2$ | 0.891 | 8.56 | 12.45 | -1.70 | 12.44 | 4.63 | FCC | [20] |
| $Al_{0.5}Cr_{0.9}FeNi_{2.5}V_{0.2}$ | 0.968 | 8.02 | 11.06 | -11.00 | 1.77 | 4.44 | FCC | [60] |
| $Al_{0.5}Cr_{9.5}Fe_{38}Mn_{9.5}Ni_{9.5}$ | 0.900 | 7.82 | 9.89 | -2.41 | 7.39 | 3.19 | FCC | [49] |
| $Al_{0.5}CrCuFeMnNi$ | 0.953 | 7.91 | 14.70 | -1.92 | 12.63 | 4.65 | BCC+FCC | [48] |
| $Al_{0.5}CrCuFeNi_2$ | 0.909 | 8.45 | 12.60 | -2.51 | 8.46 | 4.18 | FCC | [41] |
| $Al_{0.5}CrFe_{1.5}MnNi_{0.5}$ | 1.043 | 7.00 | 12.66 | -7.26 | 3.00 | 5.15 | BCC+FCC+IM | [62] |
| $Al_{0.6}CrCuFeNi_2$ | 0.927 | 8.36 | 12.72 | -3.27 | 6.51 | 4.47 | FCC | [20] |
| $Al_{0.7}CrCuFeNi_2$ | 0.944 | 8.26 | 12.81 | -3.96 | 5.36 | 4.73 | FCC | [41] |
| $Al_{0.8}Cr_{9.2}Fe_{36.8}Mn_{9.2}Ni_{9.2}$ | 0.905 | 7.80 | 10.02 | -2.65 | 6.81 | 3.33 | BCC+FCC | [49] |
| $Al_{0.8}CrCu_{1.5}FeMnNi$ | 0.971 | 7.92 | 14.74 | -1.74 | 13.44 | 4.94 | BCC+FCC | [48] |
| $Al_{0.8}CrCuFe_{1.5}MnNi$ | 0.990 | 7.68 | 14.74 | -3.31 | 7.25 | 5.07 | BCC+FCC | [48] |
| $Al_{0.8}CrCuFeMn_{1.5}Ni$ | 0.998 | 7.60 | 14.74 | -4.23 | 5.58 | 5.06 | BCC+FCC | [48] |
| $Al_{0.8}CrCuFeMnNi$ | 1.002 | 7.66 | 14.87 | -3.97 | 6.03 | 5.14 | BCC+FCC | [48] |
| $Al_{0.8}CrCuFeNi_2$ | 0.960 | 8.17 | 12.88 | -4.61 | 4.60 | 4.94 | FCC | [20] |
| $Al_{0.8}CrFe_{1.5}MnNi_{0.5}$ | 1.097 | 6.75 | 12.90 | -9.32 | 2.31 | 5.62 | BCC | [44] |
| $Al_{0.9}Cr_{0.9}Fe_{2.1}Ni_{2.1}$ | 1.008 | 7.65 | 10.84 | -10.49 | 1.76 | 5.24 | BCC+FCC | [53] |
| $Al_{0.9}CrCuFeNi_2$ | 0.976 | 8.08 | 12.94 | -5.22 | 4.05 | 5.14 | BCC+FCC | [41] |
| $Al_{1.0}Cr_9Fe_{36}Mn_9Ni_9$ | 0.908 | 7.78 | 10.11 | -2.81 | 6.46 | 3.41 | BCC+FCC | [49] |
| $Al_{1.2}CrCuFe$ | 1.165 | 6.81 | 11.50 | -0.54 | 32.54 | 6.16 | BCC+FCC | [20] |
| $Al_{1.2}CrCuFeNi_2$ | 1.021 | 7.84 | 13.02 | -6.78 | 3.07 | 5.59 | BCC+FCC | [20] |
| $Al_{1.2}CrFe_{1.5}MnNi_{0.5}$ | 1.159 | 6.46 | 12.94 | -11.29 | 1.85 | 6.02 | BCC | [44] |
| $Al_{1.5}CrCuFeNi_2$ | 1.061 | 7.62 | 13.01 | -8.05 | 2.54 | 5.92 | BCC+FCC | [41] |
| $Al_{1.8}CrCuFeNi_2$ | 1.098 | 7.41 | 12.95 | -9.07 | 2.20 | 6.16 | BCC | [41] |
| $Al_{10}Cr_9Fe_{36}Mn_{36}Ni_9$ | 1.011 | 7.14 | 11.63 | -6.87 | 2.79 | 5.09 | BCC+FCC | [46] |
| $Al_{2.0}CrCuFeNi_2$ | 1.121 | 7.29 | 12.89 | -9.63 | 2.04 | 6.29 | BCC | [41] |
| $Al_{2.2}CrCuFeNi_2$ | 1.143 | 7.17 | 12.81 | -10.12 | 1.91 | 6.40 | BCC | [41] |
| $Al_{2.5}CrCuFeNi_2$ | 1.173 | 7.00 | 12.68 | -10.74 | 1.75 | 6.51 | BCC | [41] |
| $Al_{7.5}Cr_{11}Fe_{40.2}Mn_{30}Ni_{11.3}$ | 0.981 | 7.33 | 11.73 | -5.82 | 3.40 | 4.87 | BCC+FCC+IM | [52] |
| $Al_7Cr_{20}Fe_{20}Ni_{53}$ | 0.913 | 8.31 | 9.70 | -8.42 | 2.05 | 3.77 | FCC | [56] |
| $AlCoCrFeNi_2$ | 1.019 | 7.67 | 12.98 | -12.00 | 1.83 | 5.39 | FCC+IM | [38] |
| $AlCr_{0.5}CuFeNiTi$ | 1.260 | 7.09 | 14.70 | -15.40 | 1.54 | 7.08 | BCC+FCC | [33] |
| $AlCr_{1.5}CuFeNiTi$ | 1.242 | 6.92 | 14.78 | -12.26 | 2.05 | 6.90 | BCC+FCC | [33] |
| $AlCr_2CuFeNiTi$ | 1.235 | 6.86 | 14.53 | -11.10 | 2.27 | 6.77 | BCC+FCC | [33] |
| $AlCr_3CuFeNiTi$ | 1.223 | 6.75 | 13.86 | -9.31 | 2.66 | 6.55 | BCC+FCC | [33] |
| $AlCrCu_{0.5}FeNi$ | 1.094 | 7.22 | 13.15 | -7.70 | 2.78 | 5.90 | BCC+FCC | [34] |
| AlCrCuFeMnNi | 1.032 | 7.50 | 14.90 | -5.11 | 4.63 | 5.37 | BCC | [48] |
| $AlCrCuFeNi_2$ | 0.992 | 8.00 | 12.98 | -5.78 | 3.65 | 5.30 | FCC | [48] |
| AlCrCuFeNiTi | 1.251 | 7.00 | 14.90 | -13.67 | 1.81 | 6.99 | BCC+FCC | [55] |
| AlCrFeMnNi | 1.115 | 6.80 | 13.38 | -12.48 | 1.75 | 5.81 | BCC+IM | [51] |
| AlCrFeNi | 1.154 | 6.75 | 11.53 | -13.25 | 1.45 | 6.25 | BCC+IM | [39] |
| $AlCrFeNi_{0.75}$ | 1.183 | 6.53 | 11.47 | -12.87 | 1.48 | 6.37 | BCC+IM | [58] |
| $Cr_{10}Fe_{40}Mn_{10}Ni_{10}$ | 0.893 | 7.86 | 9.59 | -2.04 | 8.51 | 2.98 | FCC | [49] |
| $Cr_{18}Fe_{27}Mn_{27}Ni_{28}$ | 0.896 | 7.93 | 11.41 | -4.24 | 4.78 | 3.67 | FCC | [15] |
| $Cr_{25}Fe_{40}Mn_{10}Ni_{25}$ | 0.904 | 7.90 | 10.72 | -3.55 | 5.60 | 2.53 | FCC | [63] |
| $Cr_2Cu_2FeMn_2Ni$ | 0.875 | 8.25 | 12.97 | 4.50 | 4.92 | 3.32 | BCC+FCC | [43] |
| $Cr_2Cu_2FeMnNi_2$ | 0.845 | 8.63 | 12.97 | 3.25 | 6.91 | 2.68 | BCC+FCC | [43] |
| $Cr_2CuFe_2Mn_2Ni_2$ | 0.885 | 8.11 | 13.15 | 0.10 | 234.16 | 3.36 | FCC | [43] |
| $Cr_2CuFe_2MnNi$ | 0.898 | 8.00 | 12.89 | 2.61 | 8.87 | 2.90 | BCC+FCC | [40] |
| $CrCu_2Fe_2Mn_2Ni$ | 0.840 | 8.50 | 12.97 | 4.69 | 4.59 | 3.38 | BCC+FCC | [43] |
| $CrCu_2Fe_2MnNi_2$ | 0.810 | 8.88 | 12.97 | 3.88 | 5.64 | 2.74 | FCC | [43] |
| $CrCuFeMn_2Ni_2$ | 0.852 | 8.43 | 12.89 | -0.49 | 44.51 | 3.55 | FCC | [43] |
| CrCuFeMnNi | 0.858 | 8.40 | 13.38 | 2.72 | 8.46 | 3.19 | BCC+FCC | [48] |
| CrCuFeMoNi | 0.976 | 8.20 | 13.38 | 4.64 | 5.75 | 3.56 | FCC | [37] |
| CrCuMnNi | 0.858 | 8.50 | 11.53 | 1.75 | 11.17 | 3.27 | BCC+FCC | [59] |
| $CrFe_{1.5}MnNi_{0.5}$ | 0.936 | 7.50 | 10.98 | -2.13 | 9.40 | 3.61 | FCC+IM | [44] |
| $CrFe_2MnNi$ | 0.906 | 7.80 | 11.08 | -3.04 | 6.59 | 3.35 | FCC | [47] |
| CrFeMnNi | 0.919 | 7.75 | 11.53 | -4.00 | 5.21 | 3.56 | FCC | [18] |
| $CrFeMnNi_2$ | 1.125 | 6.50 | 11.53 | -2.25 | 9.85 | 3.55 | BCC | [50] |
| CrFeMnNiTi | 1.189 | 7.00 | 13.38 | -13.28 | 1.85 | 6.57 | BCC+FCC+IM | [61] |
| CuFeMnNi | 0.787 | 9.00 | 11.53 | 2.75 | 6.72 | 3.41 | FCC | [42] |
| $Fe_{40}Mn_{25}Cr_{20}Ni_{15}$ | 0.918 | 7.65 | 10.97 | -2.44 | 8.09 | 3.62 | BCC+FCC | [45] |
| $Fe_{50}Mn_{27}Ni_{10}Cr_{13}$ | 0.908 | 7.67 | 9.94 | -1.61 | 10.96 | 3.73 | FCC | [57] |
| VCrFeMn | 0.878 | 8.20 | 11.08 | -5.28 | 3.76 | 3.30 | FCC | [50] |